\documentclass[aps,prr,twocolumn,groupedaddress,amsmath,amssymb,amsfonts]{revtex4-2}
\usepackage{txfonts}
\usepackage{bm} 
\usepackage{subfigure}
\usepackage{mathrsfs}
\usepackage{array}
\usepackage{amsmath,latexsym}
\usepackage{amsfonts}
\usepackage{amssymb} 
\usepackage{notes2bib}
\usepackage{braket}
\usepackage{comment} 
\usepackage{graphicx, color}
\usepackage[colorlinks=true, bookmarks=true,
bookmarksnumbered=true, bookmarkstype=toc, linkcolor=magenta,
urlcolor=cyan, citecolor=cyan]{hyperref}
\usepackage{cleveref}
\begin{document}
\title{Mean field analysis of reverse annealing for code-division multiple-access multiuser detection}
\author{Shunta Arai$^{1,2}$}
\email[]{shunta.arai.s6@dc.tohoku.ac.jp}
\author{Masayuki Ohzeki$^{1,2,3}$}
\author{Kazuyuki Tanaka$^1$}
\affiliation{$^1$Graduate School of Information Sciences, Tohoku University, Sendai 980-8579, Japan\\
$^2$Sigma-i Co., Ltd., Tokyo, Japan,\\
$^3$Institute of Innovative Research, Tokyo Institute of Technology, Nagatsuta-cho 4259, Midori-ku, Yokohama, Kanagawa 226-8503, Japan}
\date{\today}

\begin{abstract}
Code-division multiple-access (CDMA)  multiuser detection is a kind of signal recovery problems. 
The main problem of the CDMA  multiuser detection is to estimate the original signal from the degraded information.
In the CDMA multiuser detection, the first-order phase transition happens. 
The first-order phase transition degrades the estimation performance.
To avoid or mitigate the first-order phase transition, we apply adiabatic reverse annealing (ARA) to the CDMA multiuser detection.
In the ARA, we introduce the initial Hamiltonian, which corresponds to the prior information of the original signal into quantum annealing (QA) formulation.
The ground state of the initial Hamiltonian is the initial candidate solution. 
By using the prior information of the original signal, the ARA enhances the performance of the QA for the CDMA multiuser detection.
We evaluate the typical ARA performance of the CDMA multiuser detection by means of statistical mechanics using the replica method.
At first, we consider the oracle cases where the initial candidate solution is randomly generated with a fixed fraction of the original signal in the initial state.
In the oracle cases, the first-order phase transition can be avoided or mitigated by ARA if we prepare for the proper initial candidate solution. 
We validate our theoretical analysis with quantum Monte Carlo simulations. 
The theoretical results to avoid the first-order phase transition are consistent with the numerical results. 
Next, we consider the practical cases where we prepare for the initial candidate solution obtained by commonly used algorithms.
We show that the practical algorithms can exceed the threshold to avoid the first-order phase transition.
Finally, we test the performance of ARA with the initial candidate solution obtained by the practical algorithm. 
In this case, the ARA can not avoid the first-order phase transition even if the initial candidate solution exceeds the threshold to avoid the first-order phase transition.
\end{abstract}
\pacs{}
\maketitle
\section{\label{sec:sec1}Introduction}
%統計力学的解析導入
The code-division multiple-access (CDMA) multiuser detection has been used in various communication systems \cite{Verdu_1998}.
The theoretical performance of CDMA multiuser detection has been analyzed by means of statistical mechanics \cite{nishimori,Tanaka2001,Tanaka2002, Yoshida2007}. 
The CDMA multiuser detection is regarded as a type of signal recovery problems, similar to compressed sensing \cite{Donoho_2006,Donoho_2009,Kabashima_2009,Ganguli_2010}.
Statistical--mechanical analyses for signal recovery problems focus on the inference of the original information from the degraded information with noise.
The noise can be physically regarded as thermal fluctuations. 
By tuning the strength of the thermal fluctuations, the original signal can be estimated from the degraded one. 

%量子ゆらぎ+統計力学の導入
In addition to thermal fluctuations, quantum fluctuations may be used to estimate the signal. 
Several studies have demonstrated that quantum fluctuations such as the transverse field do not necessarily improve the performance of the inferences for image restoration, Sourlas codes, and CDMA \cite{inoue_2002, otsubo_2012,otsubo_2014}. The optimal decoding performance with quantum fluctuations is inferior to that with thermal fluctuations in Bayes optimal cases. However, in certain non-Bayes optimal cases; for example,
where a lower temperature than the true noise level is set, the decoding performance with finite quantum fluctuations and thermal fluctuations is superior to that with only thermal fluctuations. 
That implies the potential of the combination of quantum and thermal fluctuations for inference problems. 
%Note that even if we introduce quantum fluctuations, we can not avoid the estimation difficulty, for example, the first-order phase transition. 

%最適化の文脈だとこうだよ
The performance of optimization algorithms with quantum fluctuations, which is known as quantum annealing (QA) \cite{Kadowaki_1998,Santoro2002,Santoro_2006,Das_2008,Morita_2008,Somma_2012}  or adiabatic quantum computation (AQC) \cite{Farhi_2001,Albash_2018}, is equal to or better than that of an optimization algorithm with thermal fluctuations \cite{Denchev_2016,Albash_2018_prx}, which is known as simulated annealing (SA) \cite{Kirkpatrick_1983}. The physical implementation of QA is realized by the quantum annealer \cite{Dwave2010a,Dwave2010b,Dwave2010c,Dwave2014a,Dwave2014b}. 
The quantum annealer has been implemented in numerous applications, such as portfolio optimizations \cite{Rosenberg_2016,Venturelli_2019}, traffic optimization \cite{Neukart2017}, item listing for E-commerce \cite{Nishimura_2019}, automated guided vehicles in factories \cite{Ohzeki_2019}, machine learning \cite{Crawford2016,Benedetti_2016,Neukart_2018,Amin_2018,Khoshaman_2018}, quantum simulation \cite{King_2018,Harris_2018,Weinberg_2020}, material design \cite{Kitai_2020} and decoding algorithm \cite{SONY2020}.

%QAの説明
%1次相転移が問題であることをのべる
In a closed system, QA begins from the ground state of the transverse field term and the transverse field strength is gradually reduced. 
Following the Schrdinger
equation, the trivial ground state evolves adiabatically into a nontrivial ground state of the target Hamiltonian, which corresponds to the solution of combinatiorial optimization problems.
The quantum adiabatic theorem guarantees a theoretically sufficient condition to obtain the ground state in QA \cite{suzuki_2005}. 
The theorem indicates that the total computational time for obtaining the ground state is characterized by the minimum energy gap between the ground state and first exited state. 
The energy gap is related to the order of the phase transition.
In the case of the  first-order phase transition, the computational time for searching the ground state increases exponentially \cite{J_rg_2008,Young_2010,J_rg_2010,J_rg_2010_prl}, 
which is the worst case of QA.  

%1次相転移の回避の方法
%いろいろなやつの総論.
%68words
To avoid the first-order phase transition, many methods are proposed for example, QA with a non-stoquastic Hamiltonian \cite{Seki_2012,Seki_2015,nishimori_2017,arai2019} , inhomogeneous driving of the transverse field \cite{Susa_2018a,Susa_2018b} , and  reverse annealing (RA)  \cite{Perdomo2011,Chancellor_2017}. 
The RA is a protocol to restart the quantum dynamics starting form the resulting state of the standard procedure of QA.
We expect that the RA leads to a closer solution to the ground state as its output. 
To assess the performance of RA, we carefully classified its implementation into two methods: $adiabatic$ $reverse$ $annealing$ (ARA) and $iterated$ $reverse$ $annealing$  (IRA).
The main difference between the ARA and IRA is to incorporate the resulting state.
One is to implement the resulting state by modification of the initial Hamiltonian and the other introduce it as the initial condition.

%特にARAとは
%117 word
In the ARA, we modify the initial Hamiltonian according the resulting state.
We assume that the resulting state is a candidate solution, which is sufficiently close to the ground state of the original problem we wish to solve.
We prepare the initial Hamiltonian in the ARA such that its ground state is the candidate solution.
The procedure of the ARA is outlined as follows:
We start from the ground state of the initial Hamiltonian.
Next, we gradually increase the effects of quantum fluctuations and search locally around the candidate solution. 
Thereafter, we gradually decrease the effects of quantum fluctuations.
When the effects of quantum fluctuations disappear, the ground state or lower energy state of the original problem can be obtained. 
The ARA is rather theoretical approach for understanding the property of the RA.
Theoretical analysis has indeed shown that the ARA can avoid  the first-order phase transition for the $p$-spin model \cite{ohkuwa_2018}.
However, this protocol has not been implemented in the current quantum annealer.

%特にIRAとは
%ダイナミクスを見る必要がある。
The procedure of the IRA is slightly different from the ARA.
The difference is that the IRA starts from a classical state, which corresponds to the candidate solution without introducing the additional Hamiltonian.
A similar protocol to the IRA is feasible in the current quantum annealer.
The performance of the IRA can be analyzed from the dynamics and it significantly depends on effect of heat bath. 
In a closed system, the IRA has not enhanced the performance of the QA \cite{Yamashiro_2019}.
In an open system, the IRA has improved the performance of the QA by incorporating the relaxation mechanisms  \cite{Passarelli_2020}.

%ARAをpracticalな問題に適用する。
%CDMAの一次転移を緩和できるのでは？というイントロ
In this study, we focus on the ARA because it dramatically enhances the performance of the QA, and its performance can be analyzed by the statistical mechanics.
To the best of our knowledge, it remains unknown whether or not the ARA is useful for certain practical problems. 
We apply the ARA to the CDMA multiuser detection, which is a representative example in signal recovery problems. 
The CDMA model is mainly characterized by the ratio of the number of users to that of the measurements, which is called the pattern ratio.
In the low-temperature regions and the intermediate pattern ratio, the CDMA model has two solutions. 
This phenomenon reveals the existence of the first-order phase transition, which degrades the estimation efficacy. 
We use the ARA to mitigate or avoid the estimation difficulty. 
In the ARA process, we set the initial Hamiltonian.
The initial Hamiltonian is interpreted as prior information of the original signal in the context of the inference problems.
We expect that the prior information of the original signal will mitigate the estimation difficulty

%具体的な推定方法
We consider the marginal posterior mode (MPM) estimation by the ARA \footnote{
Originally, the ARA is applied when we search the ground state in the target Hamiltonian.
The search of the ground state corresponds to the maximum a posteriori (MAP) estimation.
The main problem of the MPM estimation is not to search the ground state but to sample the low energy state from the Gibbs--Boltzmann distribution. 
Strictly speaking, we should not utilize the term ``annealing'' because we do not perform that in this paper. 
Since the MPM estimation in the zero-temperature limit corresponds to the MAP estimation,
we utilize the term ``ARA''. 
}.
The estimated signal corresponds to the expectation of the signal over the Gibbs--Boltzmann distribution. The MPM estimation can be performed in the current quantum annealer, which provides samples from the density matrix incorporating both thermal and quantum fluctuations in about $20 \mu s$ \cite{Amin2015,Chancellor2016}.
We analyze the average MPM estimation performance with ARA at a finite temperature using the replica method. 
The MPM estimation with the ARA is regarded as the MPM estimation with quantum fluctuation incorporating the prior information of the original signal.
The typical performance of the MPM estimation with quantum fluctuations for the CDMA multiuser detection has been analyzed in the previous study \cite{otsubo_2014}. 
The connection between the present and previous studies is presented in Sec. \ref{sec:sec2a}.

In the ARA, we need to prepare for the initial candidate solution. 
In the previous study \cite{ohkuwa_2018,Yamashiro_2019}, they have not cared about how to prepare for the initial candidate solution.
We investigate whether or not  we can prepare for the proper initial candidate solution  to avoid the first-order phase transition with commonly used algorithms. 
We test the performance of the ARA with the initial candidate solution obtained by these practical algorithms.
Although the implementation of the ARA in the current quantum annealer has not yet been realized, our results provide the first theoretical demonstration of the ARA as a practical technique for signal recovery problems.

%論文の流れ
The remainder of this paper is organized as follows. 
In Section \ref{sec:sec2}, we review the previous study and present the formulation of the CDMA model with quantum fluctuations.
In Section \ref{sec:sec3}, we extend the formulation for the ARA. 
We derive the free energy under the replica symmetry (RS) ansatz and the static approximation.
In Section \ref{sec:sec4}, we illustrate the phase diagrams in the ARA. 
At first, we consider oracle cases where the initial candidate solution is randomly generated from the probability distribution given the fraction of the original signal in the initial state.
To verify the RS solutions, we perform quantum Monte Carlo simulations. 
Next, we check whether or not  we can prepare for the proper initial candidate solution  to avoid the first-order phase transition with commonly used algorithms. 
Finally, we test the performance of the ARA with the initial candidate solution attained from these practical algorithms.
 In Section \ref{sec:sec5}, we conclude the study and discuss the future research directions.

\section{\label{sec:sec2}CDMA model with quantum fluctuations}
At first, we review the previous study \cite{otsubo_2014} and show its relationship with the MPM estimation with the ARA in Sec.\ref{sec:sec2a}. 
Next, we formulate the classical CDMA model and move onto the quantum system in Sec.\ref{sec:sec2b}. 

\subsection{\label{sec:sec2a}Related Work}
The previous study \cite{otsubo_2014} analysed the performance of the MPM estimation for the CDMA model with quantum fluctuations under the standard protocol of QA. 
In particular, they shed light on the difference between quantum and thermal fluctuations.
In other words, they compared the performance between SA and QA.
In the case by SA, one controls the strength of thermal fluctuation through a parameter of temperature.
Depending on the noise in the received signal, they found the optimal strength of the thermal fluctuation known as the Nishimori temperature \cite{nishimori_1993} to retrieve the original signal in the context of the CDMA.
In the previous study, they investigated the existence of the optimal strength of the quantum fluctuation similarly to the case with thermal fluctuation.
They showed that the MPM estimation with quantum fluctuations could partially improve its performance compared to the case without quantum fluctuation. 
However, the MPM estimation with quantum fluctuations did not archive the optimal MPM performance found in the case only with thermal fluctuations.
In this sense, the thermal fluctuation is superior to the quantum fluctuation in the retrieval of the original signal of the CDMA.
Nevertheless, one of the crucial bottlenecks of the protocol in both of SA and QA to retrieved the original signal still exists.
There is the first-order phase transition in the case with the intermediate pattern ratio in the low-temperature region.
Here the temperature is a control parameter of the MPM estimation.
The existence of the first-order phase transition hampers efficient retrieval of the original signal and needs the long computation time of its execution.
As shown in the previous study, quantum fluctuation could not avoid nor mitigate the first-order phase transition. 
We thus investigate the potential of ARA, which is slightly different from the standard protocol of QA, in the present study.
In this sense, our study is placed in position as an extension of the MPM estimation with quantum fluctuations by using a different protocol of the standard QA.

\subsection{\label{sec:sec2b}Formulation}
 The main concept of the CDMA model is as follows: The digital signal of each user is modulated and transmitted to a base station through fully synchoronous channels. 
By demodulating the received signal composed of the multiuser signals and noises, we infer the original signal from the provided information. 
The following formulation is mainly based on the previous study of the CDMA model with quantum fluctuations \cite{otsubo_2014}. 
They add the transverse field to the original CDMA model and compute the partition function following the prescription of the statistical mechanics.
They used the Suzuki-Trotter decomposition to deal with quantum fluctuation written in the transverse field and the replica method to compute the averaged free energy over the quenched randomness related to the signals and modulation.
In the present study, we employ the same methods to tackle the MPM estimation of CDMA by using the ARA and setting the initial Hamiltonian depending on the initial candidate solution.

We consider that $N$ users communicate via fully synchronous channels. At the base station, the receiver obtains the signal as follows:
\begin{align}
y^\mu=\frac{1}{\sqrt{N}}\sum_{i=1}^N\eta_i^\mu \xi_i+\epsilon^\mu,
\label{E1}
\end{align}
where $\xi_i\in \{\pm1\}$, ($i=1, \dots, N$) is the original information and $\eta_i^\mu\in \{\pm1\}$ ($i=1, \dots, N$ , $\mu=1,\dots , K$)  is the spreading code for each user $i$. 
The length of the spreading codes for each user $i$ is represented by $K$. 
The channel noise $\epsilon^\mu$ is added into the received signal.
The received signal \eqref{E1} can be expressed as 
\begin{align}
\bm{y}&=\frac{1}{\sqrt{N}}\bm{\eta}\bm{\xi}+\bm{\epsilon},\label{E2}
\end{align}
for which the following notations are used:
\begin{align}
\bm{y}=\left(y^1,\dots,y^K\right)^T ,\hspace{1mm} &\bm{\xi}=\left(\xi_1,\dots,\xi_N\right)^T, \bm{\epsilon}=\left(\epsilon^1,\dots,\epsilon^K\right)^T,\label{E3}\\
\bm{\eta}&=\left(\begin{array}{cccc}
\eta_{1}^1 & \eta_{2}^1& \cdots & \eta_{N}^1 \\
\eta_{1}^2 & \eta_{2}^2& \cdots & \eta_{N}^2 \\
\vdots & \vdots & \ddots & \vdots \\
\eta_{1}^K & \eta_{2}^K& \cdots & \eta_{N}^K
\end{array}\right). \label{E4}
\end{align}  
We assume that the spreading codes and original signal are independently generated from the uniform distribution: 
\begin{align}
P(\bm{\eta})=\frac{1}{2^{NK}}
\label{E5},\\
P(\bm{\xi})=\frac{1}{2^N}.\label{E6}
\end{align}
We consider the Gaussian channels and $\epsilon^k$ is independently generated from the Gaussian distribution as follows: 
\begin{align}
P(\bm{\epsilon})=P(\bm{y}|\bm{\xi})&=\left(\frac{1}{\sqrt{2\pi T_0}}\right)^K\exp\left\{-\frac{\left|\left|\bm{\epsilon}\right|\right|_2^2}{2T_0}\right\}
\nonumber\\
&=\left(\sqrt{\frac{\beta_0}{2\pi}}\right)^K\exp\left\{-\frac{\beta_0}{2}\left|\left|\bm{y}-\frac{\bm{\eta}\bm{\xi}}{\sqrt{N}}\right|\right|_2^2\right\},\label{E7}
\end{align}
where $T_0=\beta_0^{-1}$ is the true noise scale. 

In the CDMA multiuser detection, we estimate the original signal from the received output signal and the spreading codes that are prepared for each user in advance. 
Because the output signal fluctuates owing to noise, we formulate this problem as Bayesian inference.
Subsequently, we introduce the posterior distribution of the estimated signal $\bm{\sigma}=\left(\sigma_1,\dots,\sigma_N\right)^T$  as 
\begin{align}
P(\bm{\sigma}|\bm{y})=\frac{P(\bm{y}|\bm{\sigma})P(\bm{\sigma})}{\mathrm{Tr}P(\bm{y}|\bm{\sigma})P(\bm{\sigma})},
\label{E8}
\end{align}
We define the likelihood as 
\begin{align}
P(\bm{y}|\bm{\sigma})=\left(\sqrt{\frac{\beta}{2\pi}}\right)^K\exp\left\{-\frac{\beta}{2}\left|\left|\bm{y}-\frac{\bm{\eta}\bm{\sigma}}{\sqrt{N}}\right|\right|_2^2\right\},
\label{E9}
\end{align}
where $\beta=1/T$ is the inverse temperature in statistical mechanics and corresponds to the estimated channel noise scale. 
If the true noise level is known, the estimation performance is the best and Bayes optimal. 
According to Eqs. \eqref{E8} and \eqref{E9}, the posterior distribution can be written using the Gibbs--Boltzmann distribution with the Hamiltonian $H\left(\bm{\sigma}\right)$, as follows: 
\begin{align}
P(\bm{\sigma}|\bm{y})&=\frac{1}{Z}\exp\left\{-\beta  \left(H(\bm{\sigma})+H_{\mathrm{init}}(\bm{\sigma})\right)\right\},\label{E10}\\
Z&=\mathrm{Tr}\exp\left\{-\beta  \left(H(\bm{\sigma})+H_{\mathrm{init}}(\bm{\sigma})\right)\right\} \label{E11},\\
H(\bm{\sigma})&=\frac{1}{2N}\sum_{i,j}\sum_{\mu=1}^K\eta_i^\mu \eta_j^\mu\sigma_i\sigma_j-\frac{1}{\sqrt{N}}\sum_{i=1}^N\sum_{\mu=1}^K\eta_i^\mu y^\mu\sigma_i,\label{E12}
\end{align}
where $Z$ is the partition function and $H_{\mathrm{init}}(\bm{\sigma})$ is the initial Hamiltonian, which represents the prior information of the estimated signal. 
We generally assume that the prior of the estimated signal follows the uniform distribution 
\begin{align}
P(\bm{\sigma})=\frac{1}{2^N}.
\label{E13}
\end{align}
In this case, we can omit the initial Hamiltonian from Eqs. \eqref{E10} and \eqref{E11}.

To estimate the original signal, we consider the MPM estimation.
The estimation performance can be evaluated by the overlap between the original and estimated signal as $\mathcal{M}=1/N \sum_{i=1}^N \xi_i \mathrm{sgn}\langle\sigma_i \rangle$,
where $\langle\cdot \rangle$ is the expectation over the posterior distribution $P(\bm{\sigma}|\bm{y})$ and $\mathrm{sgn}(\cdot)$ is the signum function.
This quantity is expected to exhibit a ``self-averaging'' property in the thermodynamics limit $N\rightarrow\infty$.
This means that the observables, such as the overlap for a quenched realization of the data $\bm{y}$, $\bm{\eta}$, and $\bm{\xi}$, are equivalent to the expectation of itself over the data distribution $P(\bm{\eta}) P(\bm{\xi}) P(\bm{y}|\bm{\xi})$. 
In this case, the overlap can be expressed as $\lim_{N\rightarrow\infty}\mathcal{M}=[\xi_i\mathrm{sgn}\langle \sigma_i\rangle]$, where the bracket $[\cdot]$ indicates the expectation over the data distribution.

It is straightforward to extend the above formulation into the quantum mechanical version: 
\begin{align}
\hat{H}&=s\hat{H}_0+(1-s)\hat{H}_{\mathrm{TF}}, \label{E14}\\
\hat{H}_0&=\frac{1}{2N}\sum_{i,j}\sum_{\mu=1}^K\eta_i^\mu \eta_j^\mu\hat{\sigma}_i^z\hat{\sigma}_j^z-\frac{1}{\sqrt{N}}\sum_{i=1}^N\sum_{\mu=1}^K\eta_i^\mu y^\mu \hat{\sigma}_i^z,\label{E15}\\
\hat{H}_{\mathrm{TF}}&=-\sum_{i=1}^N\hat{\sigma}_i^x\label{E16},
\end{align}
where $\hat{\sigma}_i^z$ and  $\hat{\sigma}_i^x$ are the $z$ and $x$ components of the Pauli matrices at site $i$, respectively. 
In this case, $\hat{H}_0$ consists of the $z$ components of the Pauli matrices and $\hat{H}_{\mathrm{TF}}$ is composed of the $x$ components of the Pauli matrices. 
We parameterize the Hamiltonian \eqref{E14} with the annealing parameter $s$ for application to the ARA. 

As in the classical case, we consider the MPM estimation with quantum fluctuations.
The posterior distribution can be written as $\hat{\rho}=\exp\left\{-\beta \hat{H}\right\}/\mathrm{Tr}\exp\left\{-\beta \hat{H}\right\}$
where $\mathrm{Tr}$ denotes the summation over all possible spin configurations in the z-basis.
The performance of the MPM estimation with quantum fluctuations can be evaluated by $\mathcal{M}=1/N\sum_{i=1}^N \xi_i \mathrm{sgn}\left(\mathrm{Tr} \hat{\sigma}_i^z\hat{\rho}\right) $.

 \section{\label{sec:sec3}Mean field analysis}
 Following Ref. \cite{ohkuwa_2018}, 
 we extend the CDMA model with quantum fluctuations \cite{otsubo_2014} to the ARA formulation as 
 \begin{align}
 \hat{H}&=s\hat{H}_0+(1-s)(1-\lambda)\hat{H}_{\mathrm{init}}+(1-s)\lambda \hat{H}_{\mathrm{TF}}\label{E17},\\
 \hat{H}_{\mathrm{init}}&=-\sum_{i=1}^N\tau_i \hat{\sigma}_i^z,\label{E18}
 \end{align}
 where $\lambda\hspace{2pt}(0\leq \lambda\leq 1)$ is the RA parameter.
 The initial candidate solution is denoted by $\tau_i=\pm1  (i=1,\dots,N)$ that 
 is expected to be close to the original signal $\xi_i$.
We introduce the probability distribution of the initial candidate solution as follows: 
 \begin{align}
P(\bm{\tau})=\prod_{i=1}^NP\left(\tau_i\right)=\prod_{i=1}^N\left(c_1\delta(\tau_i-\xi_i)+c_{-1}\delta(\tau_i+\xi_i)\right),
\label{E19}
 \end{align}
 where we define $c_1=c$ and $c_{-1}=1-c$.
 The number $c\hspace{2pt} (0\leq c\leq 1)$ denotes the fraction of the original signal $\tau_i=\xi_i$ in the initial state as 
\begin{align}
c=\frac{1}{N}\sum_{i=1}^N\delta_{\tau_i,\xi_i}%=\frac{\mathcal{M}+1}{2}
\label{E20}.
\end{align}
 The prior information can be incorporated through Eq.\eqref{E19}.
 The main concept of the MPM estimation with the ARA is to avoid or mitigate the first-order phase transition by controlling the RA parameter and utilizing the prior information.

The typical behaviors of the order parameters such as the overlap can be obtained via the free energy. 
We calculate the partition function $Z=\mathrm{Tr}\exp\left(-\beta \hat{H}\right)$ and derive the RS free energy in the limit of $N,K\rightarrow \infty$, while maintaining the pattern ratio $\alpha\equiv K/N=O(1)$. 
The free energy density $f$ can be evaluated as $-\beta f=\lim_{N\rightarrow \infty} (1/N)[\ln Z]$ where $[\cdot]$ denotes the configuration average over the data distribution $P(\bm{y}|\bm{\xi})P(\bm{\eta}) P(\bm{\xi})P(\bm{\tau})$.
When computing $f$, we have two difficulties. 
The first one is the non-commutativity of the spin operator from Eq.\eqref{E16}.
We can not apply the mean-field analysis directly into the partition function of Eq.\eqref{E17}.
The second one is to compute  $[\ln Z]$. 
In general, it is difficult to directly evaluate  $[\ln Z]$. 
We remove two difficulties by using two techniques.

Firstly, to exclude the non-commutativity  of the spin operator, we employ the Suzuki--Trotter (ST) decomposition \cite{suzuki_1976} in the partition function:
\begin{align}
Z&=\lim_{M\rightarrow \infty}\mathrm{Tr}\left\{\exp\left(-\frac{\beta}{M}\left(\hat{H}_0+\hat{H}_{\mathrm{init}}\right)\right)\exp\left(-\frac{\beta}{M}\hat{H}_{\mathrm{TF}}\right)\right\}^M\nonumber\\
&=\lim_{M\rightarrow \infty}Z_M,\label{E21}
\end{align}
where 
\begin{widetext}
\begin{align}
Z_M=&\mathrm{Tr}\prod_{t=1}^M\left(\frac{1}{2}\sinh\left(\frac{2\beta (1-s)\lambda}{M}\right)\right)^{\frac{N}{2}}\exp\left\{-\frac{\beta s}{2NM}\sum_{i,j}\sum_{\mu=1}^K \eta_i^\mu \eta_j^\mu \sigma_{i}(t)\sigma_{j}(t)+\frac{\beta s}{M\sqrt{N}}\sum_{i=1}^N\sum_{\mu =1}^K\eta_i^\mu y^\mu \sigma_{i}(t)\right.\nonumber\\
&\left.+\frac{\beta(1-s)(1-\lambda)}{M}\sum_{i=1}^N\tau_i\sigma_{i}(t)+\frac{1}{2}\ln \coth\left(\frac{\beta(1-s)\lambda}{M}\right)\sum_{i=1}^N\sigma_{i}(t)\sigma_{i}(t+1)\right\}
\label{E22}
\end{align}
in which the symbol $t$ is the index of the Trotter slice and $M$ is the Trotter number. 
We impose the periodic boundary conditions, $\sigma_i(M+1)=\sigma_i(1)$ for all $i$.
By using the ST decomposition, we can map the quantum system into the identical classical system.
The difficulty from the non-commutativity  of the spin operator is removed. 
Above expressions, we replace $\sigma_i^z(t)$ with the classical spin $\sigma_i(t)\in\{ -1,+1\}$.
In this case, the symbol $\mathrm{Tr}$ represents the trace over the classical spins.
The $x$ component of the Pauli matrix yields the last term in Eq.\eqref{E22}.

Secondly, to evaluate $[\ln Z]$, we exploit the replica method \cite{replica_method}:
$[\ln Z]=\lim_{n\rightarrow0}([Z^n]-1)/n$.
The symbol $n$ denotes the number of replicas.
By using the replica method, we can take the configuration average for the replicated partition function $Z^n$ and the limit of $n\rightarrow 0$.
When manipulating the configuration average over $P(\bm{y}|\bm{\xi})P(\bm{\eta}) P(\bm{\xi})P(\bm{\tau})$, we introduce the order parameters and their conjugate parameters through the delta function and its Fourier integral representation as follows: the magnetization $m_a(t)=(1/N)\sum_{i=1}^N\xi_i\sigma_{ia}(t)$, the spin glass order parameter $q_{ab}(t,t')=(1/N)\sum_{i=1}^N\sigma_{ia}(t)\sigma_{ib}(t')$ $(a\neq b)$, and the correlation between each Trotter slice $R_{a}(t,t')=(1/N)\sum_{i=1}^N\sigma_{ia}(t)\sigma_{ia}(t')$. The conjugate parameters are denoted by  $\tilde{m}_a(t), \tilde{q}_{ab}(t,t')$  $(a\neq b)$ ,and $\tilde{R}_{a}(t,t')$.
These conjugate parameters appear in manipulation of several integrals over order parameters to compute the partition function as detailed in Supplemental Material \cite{supp1}
The symbols $a$ and $b$ represent the replica indices.
Under the RS ansatz and static approximation: $
 m_a(t)=m, q_{ab}(t,t')=q,R_{a}(t,t') =R,\tilde{m}_a(t)=\tilde{m}, \tilde{q}_{ab}(t,t')=\tilde{q}, \tilde{R}_{a}(t,t') = \tilde{R}$, 
we can finally obtain the RS free energy density: 
\begin{align}
-\beta f_{\mathrm{RS}}&=\underset{\substack{m,q,R\\ \tilde{m},\tilde{q},\tilde{R}}}{\mathrm{extr}}\left[\frac{\alpha}{2}\left\{-\ln (1+\beta s(R-q))+\beta s\left((R-1)+ \frac{1+\beta_0}{\beta_0}+\frac{2m-q-(1+\beta_0^{-1})}{1+\beta s (R-q)}\right)\right\}\right.\nonumber\\
&\left.-m\tilde{m}-R\tilde{R}+\frac{1}{2}q\tilde{q}+\sum_{a=\pm1}c_a\int Dz\ln \int Dy 2\cosh\sqrt{g_{a}^2+(\beta(1-s)\lambda)^2}\right],\label{E23}
\end{align}
where
\begin{align}
g_{a}=\tilde{m}+a\beta(1-\lambda)(1-s)+\sqrt{\tilde{q}}z+\sqrt{2\tilde{R}-\tilde{q}}y,\label{E24}
\end{align}
in which $Dz$ means that the Gaussian measure $Dz:=dz/\sqrt{2\pi}  e^{-z^2/2}$ and $Dy$ is the same as $Dz$.
Here $\mathrm{extr}$ represents the extremization by changing the order parameters $m,q$ and $R$, and their conjugate parameters as $\tilde{m},\tilde{q}$ and $\tilde{R}$.
The extremum point is determined by the saddle-point conditions and characterizes the free energy density.
The expression in Eq.\eqref{E23} for $\lambda=1$ can be reduced to the RS free energy density derived in \cite{otsubo_2014}.
The detailed derivation of Eq. \eqref{E23} is written in Supplemental Material\cite{supp1}.
The saddle-point equations are referred in Appendix \ref{appendix_a}. 
Below we investigate the phase transition of the order parameters while tuning the external parameters as the strength of the transverse field, the pattern ratio etc..
Then we numerically solve the saddle-point equations for each set of external parameters.
\begin{figure*}[t]
\subfigure[\label{fig:fig_1a}]{\includegraphics[width=58mm]{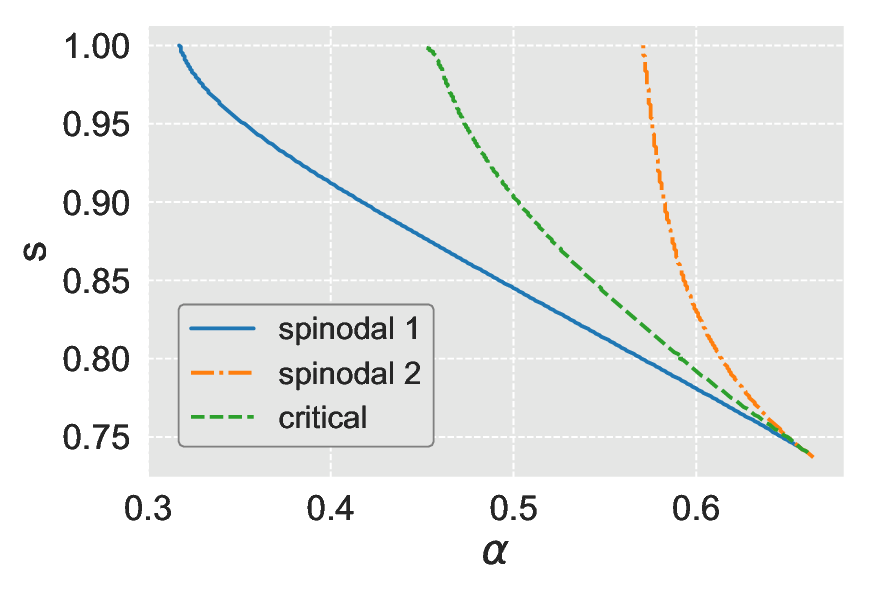}}
\subfigure[\label{fig:fig_1b}]{\includegraphics[width=58mm]{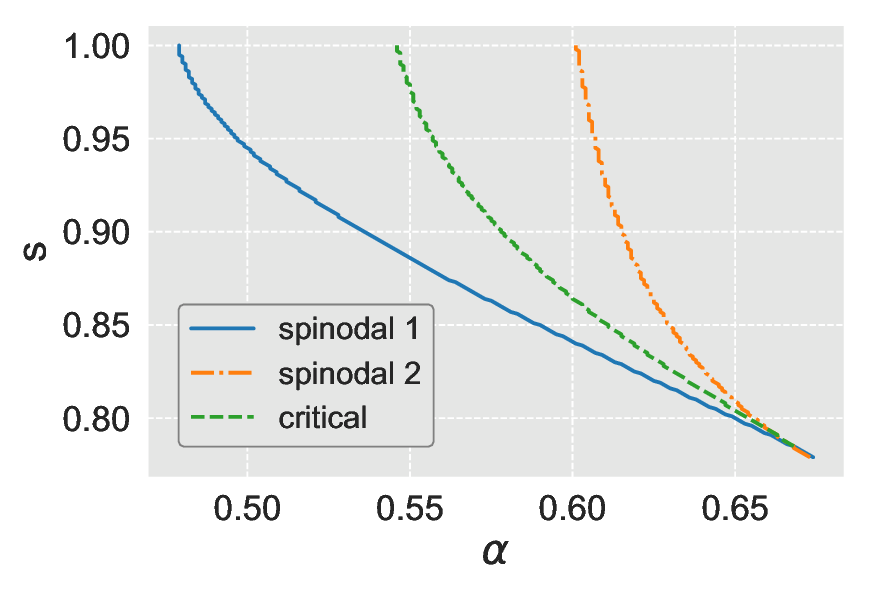}}
\subfigure[\label{fig:fig_1c}]{\includegraphics[width=58mm]{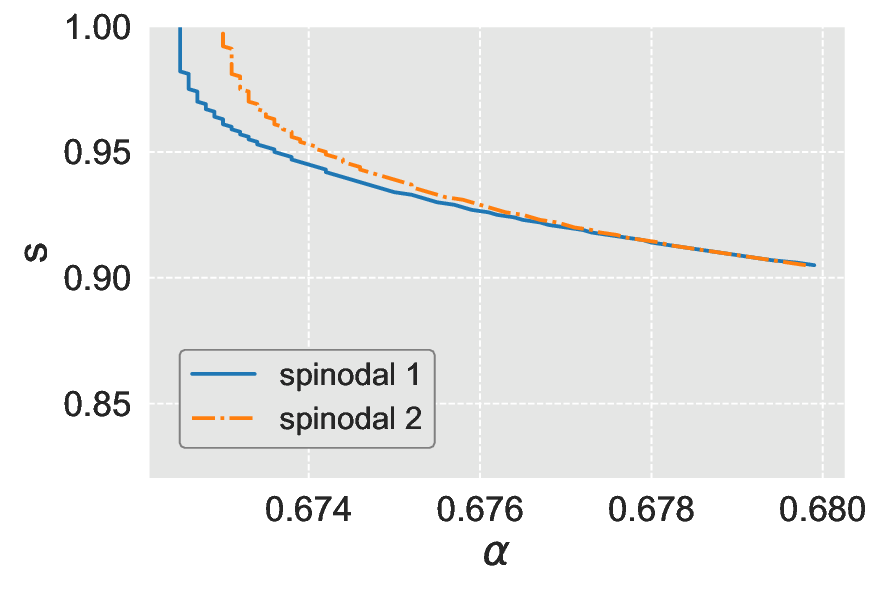}}

\caption{Phase diagram of CDMA model in the ARA with $\lambda=1$. 
The horizontal axis denotes the pattern ratio.
The vertical axis represents the annealing parameter.
The experimental settings are (a) $T_0=0$, (b) $T_0=0.05$, and (c) $T_0=0.1$. The ``spinodal 1'' and ``spinodal 2'' curves indicate the solutions from the two different branches. The ``critical'' curve denotes the point at which the RS free energy takes the same value.}
\label{fig:fig_1}
\end{figure*}
\begin{comment}
\begin{align}
-\beta f_{\mathrm{RS}}&=\alpha\left\{-\frac{1}{2}\ln (1+\beta s(R-q))+\frac{\beta s}{2}(R-1)+\frac{\beta s}{2}\left( \frac{1+\beta_0}{\beta_0}+\frac{2m-q-(1+\beta_0^{-1})}{1+\beta s (R-q)}\right)\right\}
\nonumber\\
&-\frac{\alpha \beta sm}{1+\beta s(R-q)}-R\tilde{R}+\frac{1}{2}\frac{\alpha \beta^2 s^2q\left(q-2m+1+\beta_0^{-1}\right)}{(1+\beta s(R-q))^2}\nonumber\\
&+c\int Dz\ln \int Dy 2\cosh \beta \sqrt{g_{+}^2+((1-s)\lambda)^2}+(1-c)\int Dz\ln \int Dy 2\cosh\beta \sqrt{g_{-}^2+((1-s)\lambda)^2},\label{E29}
\end{align}
where
\begin{align}
g_{\pm}=\frac{\alpha s}{1+\beta s(R-q)}\pm(1-\lambda)(1-s)+s\sqrt{\frac{\alpha \left(q-2m+1+\beta_0^{-1}\right)}{(1+\beta s(R-q))^2}}z+ s\sqrt{\frac{\alpha(R-q)}{1+\beta s(R-q)}}y,\label{E30}
\end{align}
\end{comment}
\end{widetext}
\section{\label{sec:sec4}Numerical results}
%概要の地図を示す. 
%3章の結果を元に4章に移る。
%4.Aでは oracle cases
%4.Bでは practical casesを扱う
In this section, we evaluate the typical performance of the ARA based on the results attained in  Section \ref{sec:sec3}.
In Section \ref{sec:sec41}, we consider the oracle cases where the initial candidate solution is randomly generated from Eq.\eqref{E19} given the fraction of the original signal in the initial state.
In Section  \ref{sec:sec42}, we consider the practical cases where we prepare for the initial candidate solution with commonly used algorithms. 
We compare the performance of the ARA with the oracle cases and the practical cases.

\subsection{\label{sec:sec41}Analysis of the ARA in oracle cases}
%Aの中でも地図を示す.
%なぜこれをするのかはsubsec毎に伏線を貼って, 次回収という流れでいく. 

%lambda=1の場合の相図の説明
We numerically solve the saddle-point equations in Eqs. \eqref{A1} to \eqref{A6} with the temperature  $T=0.1$.
We set the several RA parameter $\lambda$.
We start from the ARA with $\lambda=1$ which corresponds to the vanilla QA \cite{otsubo_2014}.
We show that the CDMA model has the first-order phase transition in the intermediate pattern ratio. 
Next, we consider the classical case with $\lambda=0$ to validate the RS ansatz without the static approximation.
Finally, we move onto the finite  $\lambda$. 
We exhibit that the ARA can avoid or mitigate the first-order phase transition if we prepare for the proper initial candidate solution. 

\begin{figure*}[t]
\subfigure[\label{fig:fig_2a}]{\includegraphics[width=54mm]{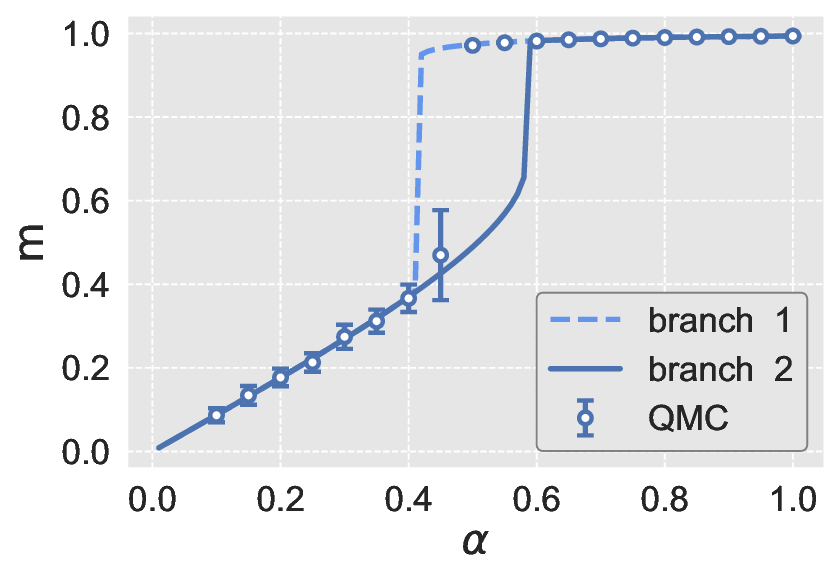}}
\subfigure[\label{fig:fig_2b}]{\includegraphics[width=54mm]{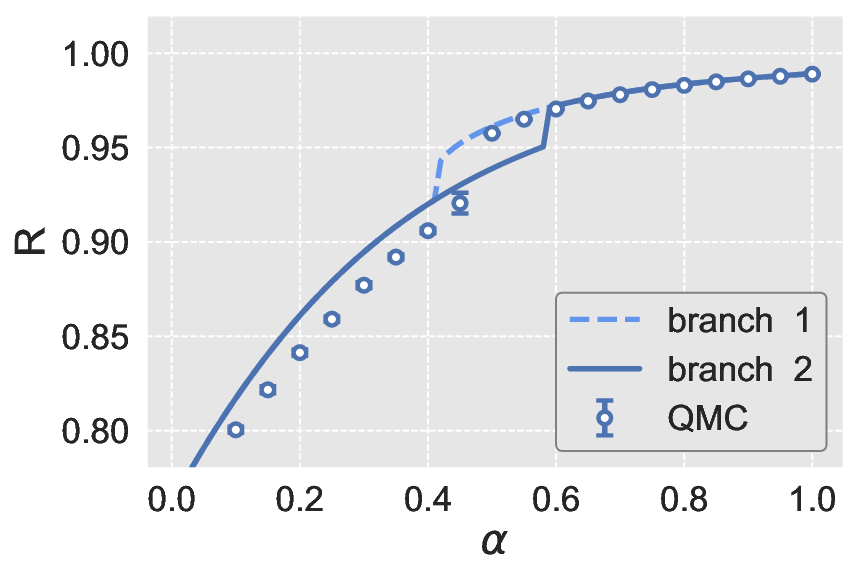}}
\caption{Dependence of the order parameters on the pattern ratio for the fixed annealing parameter $s=0.9$. The vertical axes denote these order parameters: (a) magnetization and (b) correlation between Trotter slices. 
The  solid blue curve and  dashed blue curve denote the two different branches that are obtained from the saddle-point equations. 
The circles represent the results obtained by the quantum Monte Carlo simulations.}
\label{fig:fig_2}
\end{figure*}
\begin{figure*}[t]
\subfigure[\label{fig:fig_3a}]{\includegraphics[width=52mm]{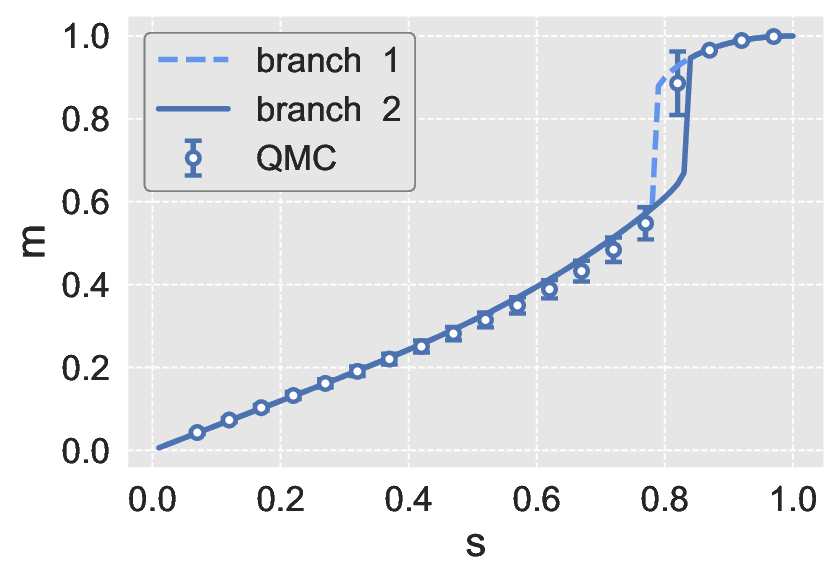}}
\subfigure[\label{fig:fig_3b}]{\includegraphics[width=52mm]{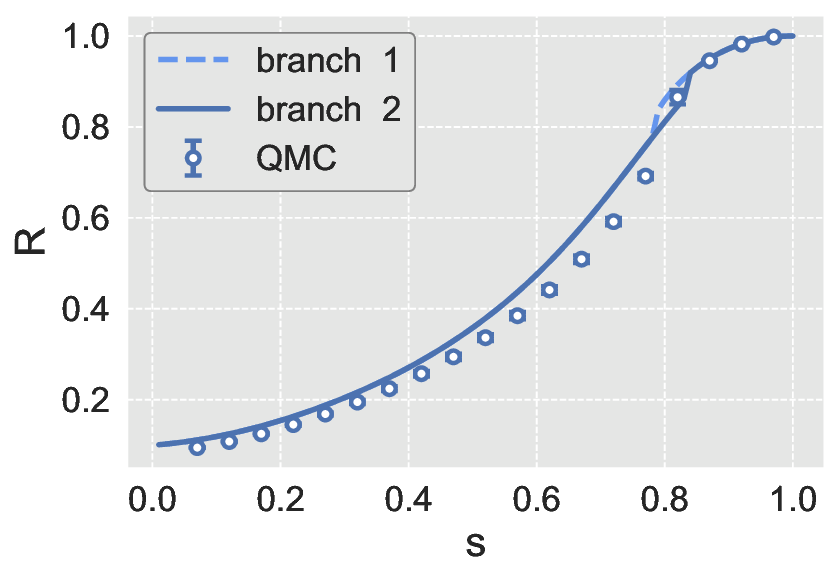}}
\caption{Dependence of order parameters on annealing parameter for fixed pattern ratio $\alpha=0.6$. The same symbols as those in Fig. \ref{fig:fig_2} are used. }
\label{fig:fig_3}
\end{figure*}

%相図概要
\subsubsection{\label{sec:sec411} ARA with $\lambda=1$}
Let us begin with the ARA with $\lambda=1$.
The phase diagrams for the true noise scale $T_0=0, 0.05$ and $0.1$ are displayed in Fig. \ref{fig:fig_1}.
The blue solid curve and orange dash-dotted curve indicate the spinodal curves where the solutions for each initial condition disappears in Figs. \ref{fig:fig_1a} - \ref{fig:fig_1c}.
Two solutions coexist between the two spinodal curves. 
From these figures, we can establish the existence of the first-order phase transition in the intermediate pattern ratio and under the weak strength of the transverse field.
The green dotted curve denotes the critical point at which the RS free energy  takes the same value.
In Fig. \ref{fig:fig_1c}, we do not write down the curve because we can not distinguish the critical point from the spinodal points in this scale.
Higher noise results in a narrower region in which the two solutions coexist.
Although the noise mitigates the first-order phase transition, it decreases the overlap between the original signal and the estimated one. 

%1次転移があるとなぜ信号推定が困難なのかの説明
Next, we consider why the first-order phase transition is troublesome in estimating the original signal. 
The difficulty of estimating the original signal is related to the free energy landscape.
We take Fig.\ref{fig:fig_1a} as an example.
On the right side of spinodal curve 2, it is easy to estimate the original signal because the free energy exhibits a minimum, which is a good estimator.
When we set the pattern ratio as $\alpha=0.6$, we encounter the first-order phase transition at $s\simeq 0.8$.
The free energy landscape has two valleys. 
At spinodal curve 2, the free energy landscape is transformed into a simple valley. 
In this case, it is comparatively easy to estimate the original signal.
For $\alpha=0.5$, the spinodal curve 2 does not exist. 
The free energy landscape maintains two valleys.
We can not efficiently estimate the original signal because the metastable state remains. 
For $\alpha=0.4$, the critical point does not exist.
In this case, we can not obtain the original signal information-theoretically.
The ground state or low energy state does not correspond to the original signal at $s=1$. 
The minima of the free energy do not provide us with an effective estimation. 

 \begin{figure*}[t]
\subfigure[\label{fig:fig_4a}]{\includegraphics[width=55mm]{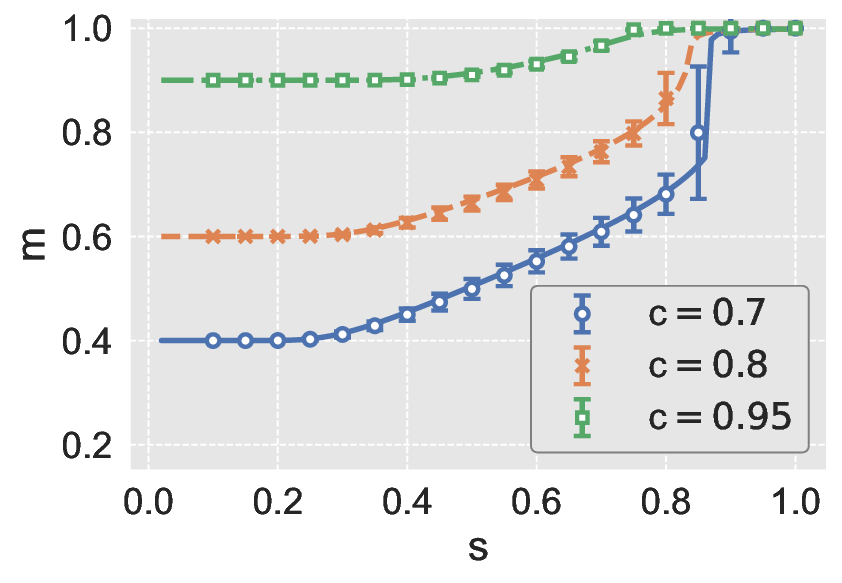}}
\subfigure[\label{fig:fig_4b}]{\includegraphics[width=55mm]{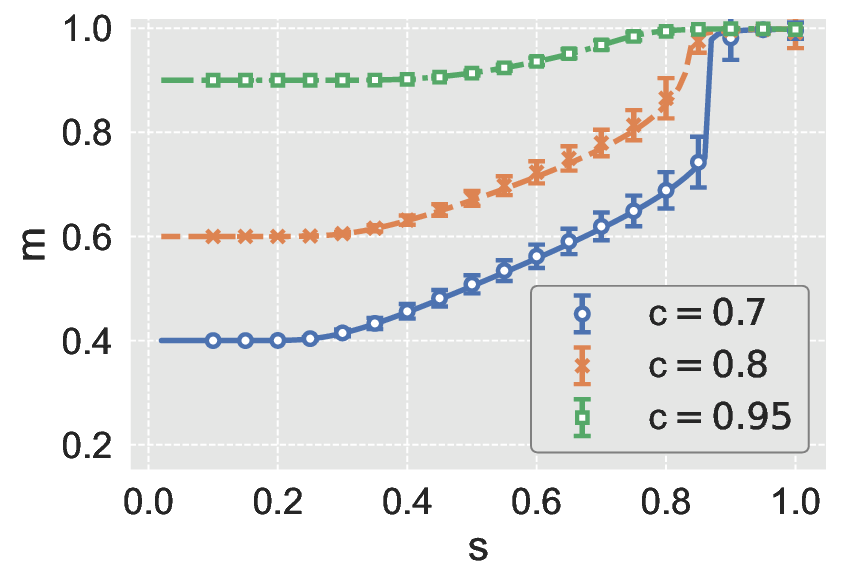}}
\caption{Dependence of magnetization on annealing parameter with $\lambda=0$. 
The experimental settings are as follows: (a) is $\alpha=0.6$ and $T_0=0$, and (b) is $\alpha=0.62$ and $T_0=0.05$. Both axes are the same as those in Fig. \ref{fig:fig_3a}. }
\label{fig:fig_4}
\end{figure*}

\begin{figure*}[t]
\subfigure[\label{fig:fig_5a}]{\includegraphics[width=55mm]{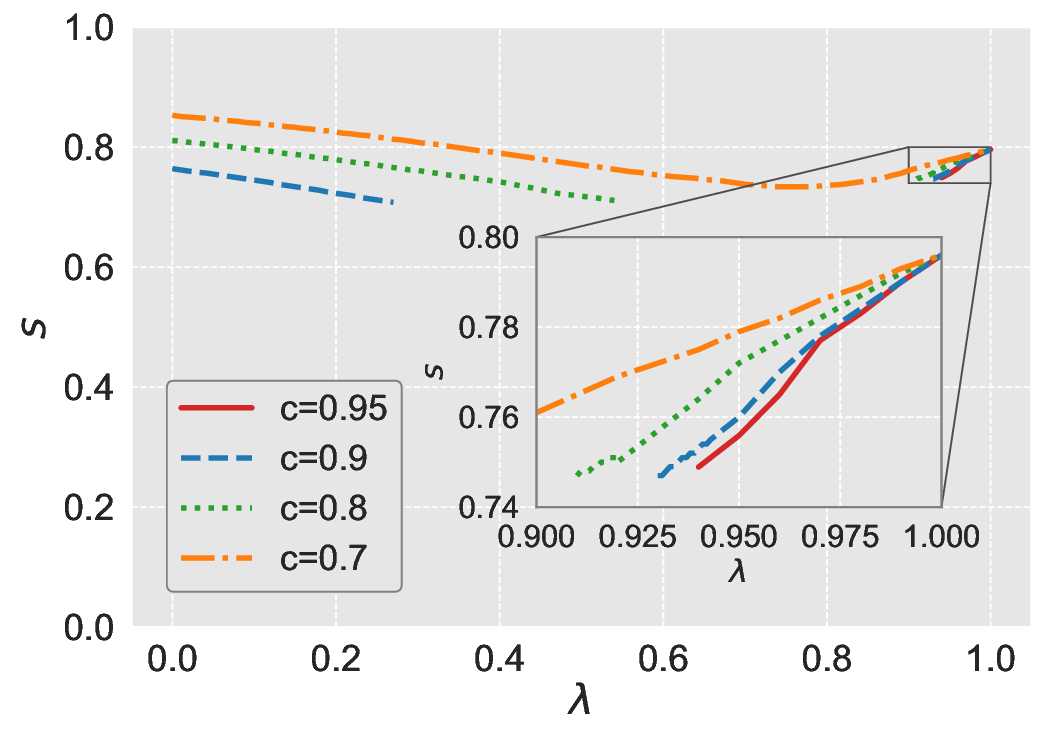}}
\subfigure[\label{fig:fig_5b}]{\includegraphics[width=55mm]{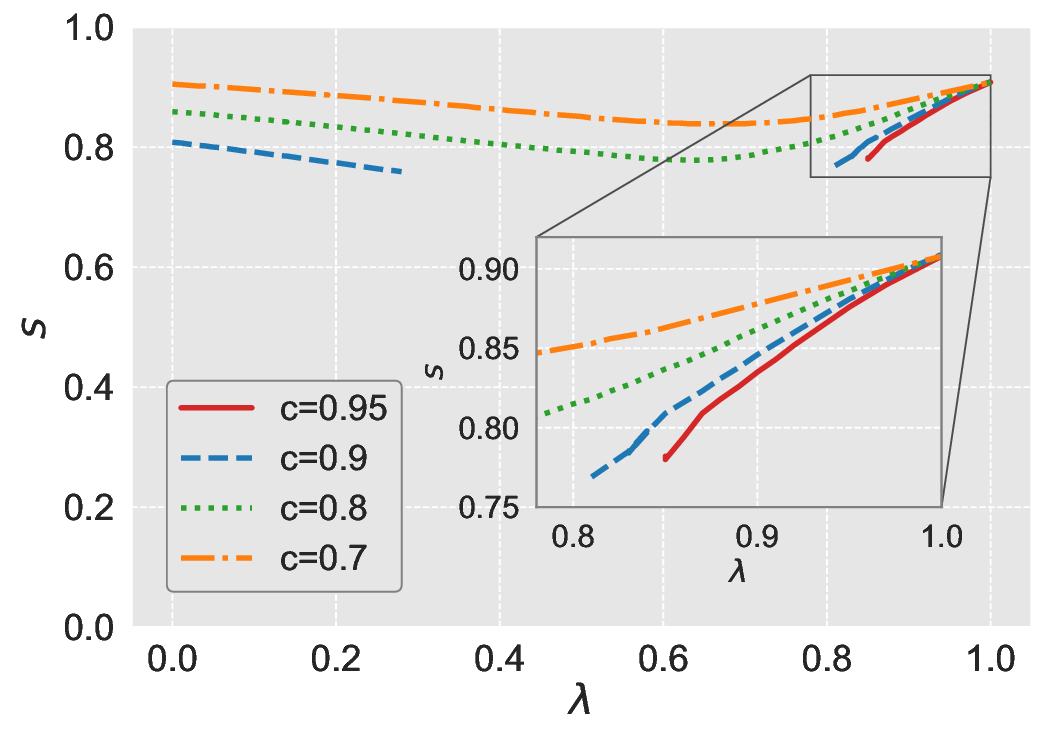}}
\caption{Phase diagrams of CDMA model in ARA for four different values of $c$. The horizontal axis denotes the RA parameter. The vertical axis denotes the annealing parameter. 
These curves represent the points at which the first-order phase transitions occur.
The experimental settings are (a) $\alpha=0.6$ and (b) $\alpha=0.5$.}
\label{fig:fig_5}
\end{figure*}

\begin{figure*}[t]
\subfigure[\label{fig:fig_6a}]{\includegraphics[width=60mm]{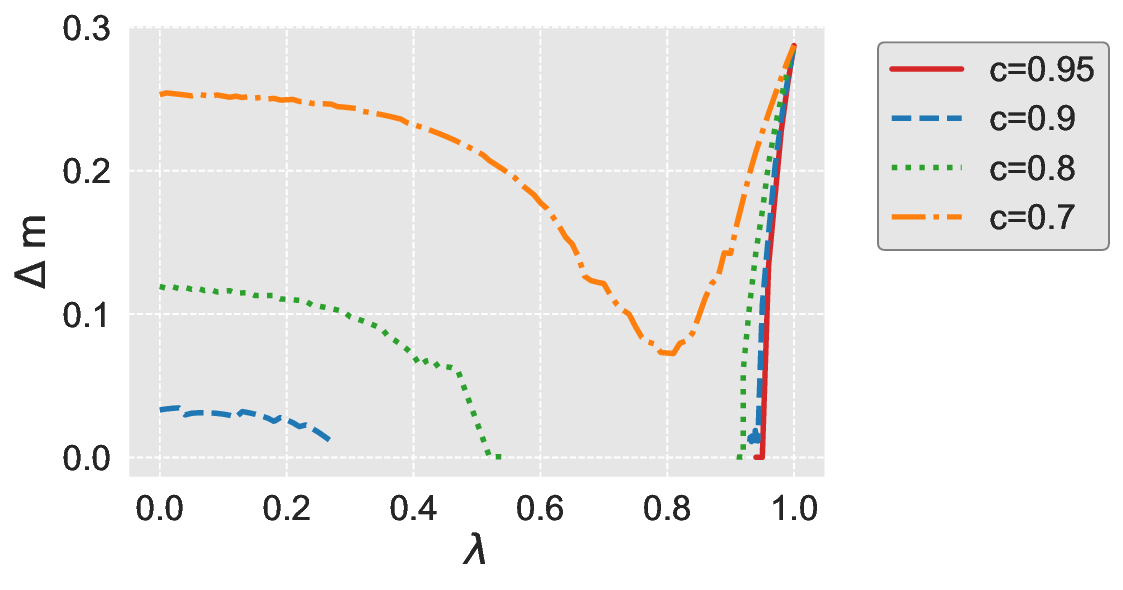}}
\subfigure[\label{fig:fig_6b}]{\includegraphics[width=60mm]{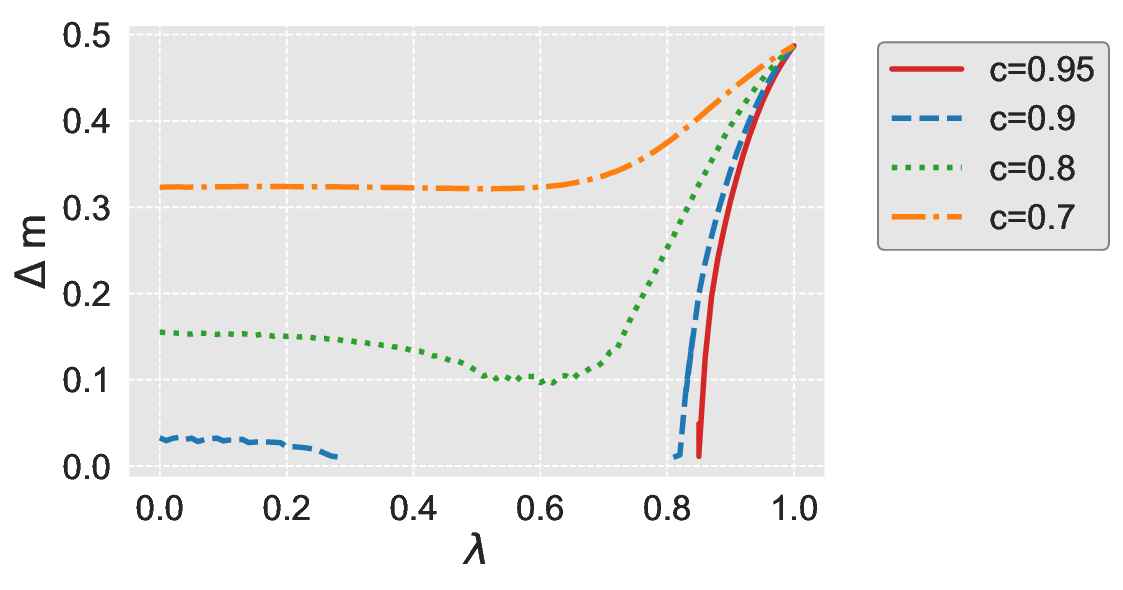}}
\caption{Differences in magnetization between two local minima at first-order phase transition in Figs. \ref{fig:fig_5a} and \ref{fig:fig_5b}. The vertical axis denotes the differences in the magnetization between the two local minima at the first-order phase transition. The horizontal axis denotes the RA parameter.
The experimental settings are (a) $\alpha=0.6$ and (b) $\alpha=0.5$. }
\label{fig:fig_6}
\end{figure*}

%lambda=1の場合のQMCによるvalicdation
To verify the RS ansatz and the static approximation, we perform quantum Monte Carlo simulations for the CDMA model.
We set the system size as $N=500$, the Trotter number as $M=50$, the temperature as $T=0.1$, and the true noise scale as $T_0=0$.
We use a $100000$ Monte Carlo step (MCS) average after $50000$ MCS equilibrations for each instance. We take the configuration average over the spreading codes and the original signals by randomly generating $50$ instances.
We plot the behavior of the order parameters with respect to the pattern ratio for the fixed annealing parameter $s=0.9$ in Fig. \ref{fig:fig_2} and the annealing parameter for the fixed pattern ratio $\alpha=0.6$ in Fig. \ref{fig:fig_3}. The error bar is given by the standard deviation.
The results obtained by the quantum Monte Carlo simulations are the averages over all of the Trotter slices. 
Following Ref. \cite{Yoshida_2006}, we adopt the magnetization to quantify the performance of the MPM estimation.
In this study, we refer to the solution representing the ``spinodal 1'' curve as ``branch 1'' and to the solution representing the ``spinodal 2'' curve as ``branch 2''.  According to Fig. \ref{fig:fig_2}, the results obtained by the quantum Monte Carlo simulations are consistent with the RS solutions, with the exception of the low pattern ratio. 
Figure.\ref{fig:fig_3} shows that the numerical results for the magnetization is consistent with the RS solutions, except for the intermediate values of the annealing parameter.
The numerical result of the correlation between the Trotter slices does not follow the RS solutions other than the large annealing parameter, which is close to  $1$
 \footnote{
We verify the dependence of the order parameters on the Trotter number in our simulations.
As we increase the Trotter number, the deviation of the correlation between the Trotter slices from the RS solutions decreases. 
The qualitative results are similar to those in the magnetization.}.
To investigate the deviations between the numerical results and the RS solutions due to the replica symmetry breaking (RSB), we compute the Almeida--Thouless (AT) condition and the entropy. 
The details of these formula are written in Appendix \ref{appendix_a}.
In these problem settings, the AT condition is not broken, and the entropy is positive. 
The deviations between the the numerical results and the RS solutions probably result from the breaking of the static approximation. 

\begin{figure*}[t]
\subfigure[\label{fig:fig_7a}]{\includegraphics[width=55mm]{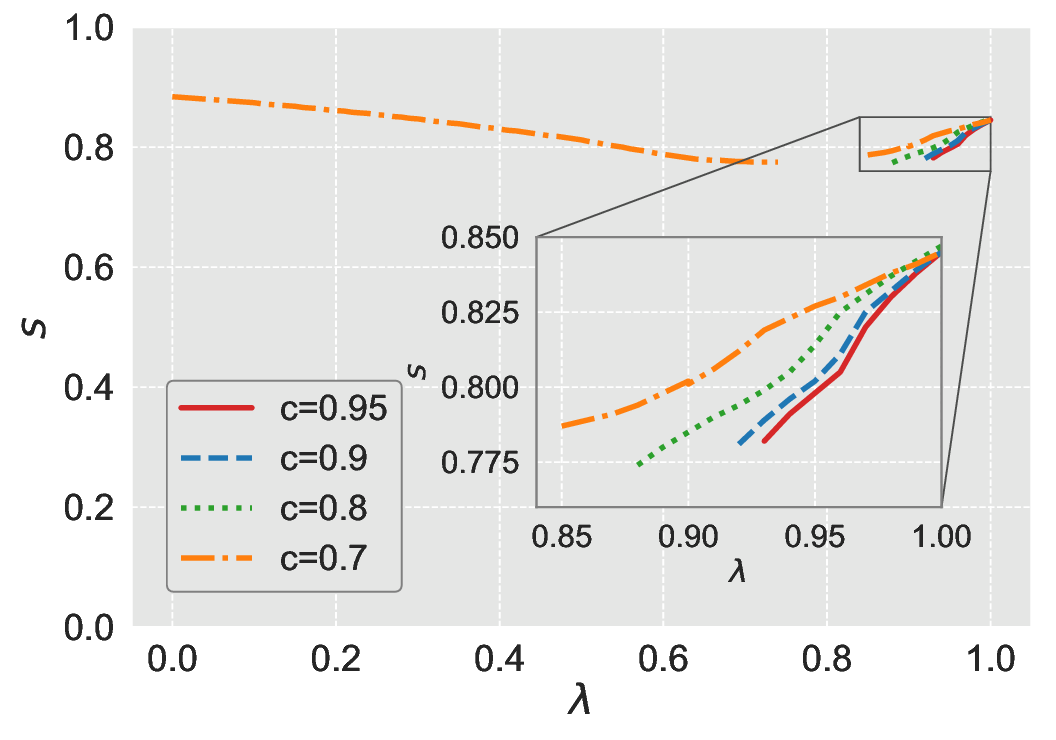}}
\subfigure[\label{fig:fig_7b}]{\includegraphics[width=55mm]{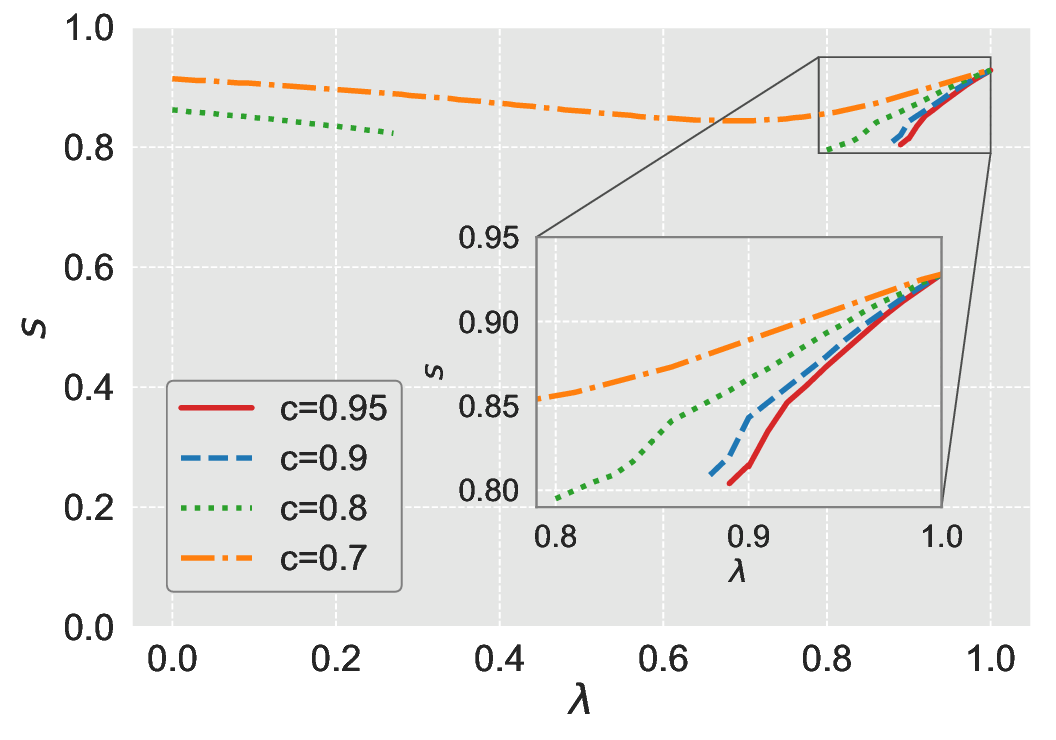}}
\caption{Phase diagrams of CDMA model in ARA for four different values of $c$. Both axes are the same as those in Fig. \ref{fig:fig_5}. The experimental settings are (a) $\alpha=0.62$ and (b) $\alpha=0.57$. }
\label{fig:fig_7}
\end{figure*}
\begin{figure*}[t]
\subfigure[\label{fig:fig_8a}]{\includegraphics[width=60mm]{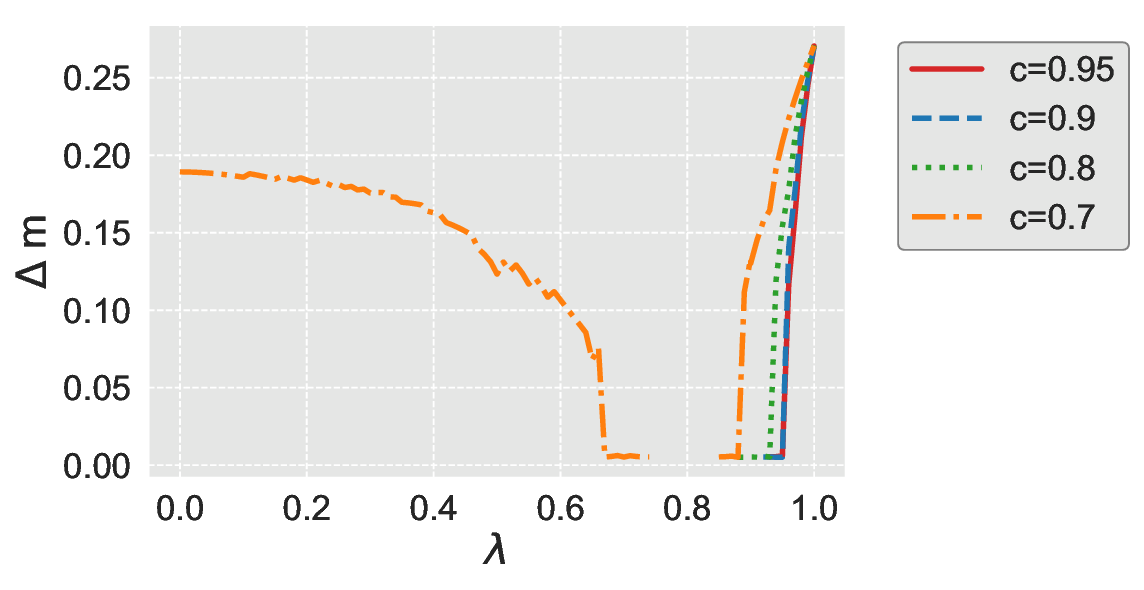}}
\subfigure[\label{fig:fig_8b}]{\includegraphics[width=60mm]{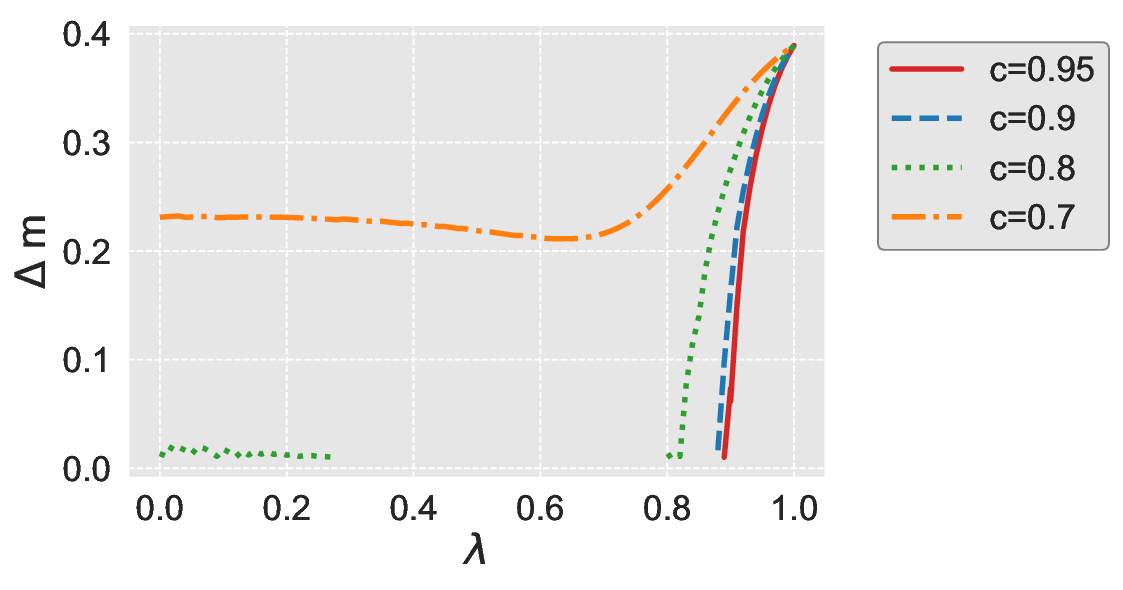}}
\caption{The differences of the magnetization between the two local minima at the first-order phase transition in Figs. \ref{fig:fig_6a} and \ref{fig:fig_6b}. Both axes are the same as those in Fig.\ref{fig:fig_6}.
The experimental settings are (a) $\alpha=0.62$ and (b) $\alpha=0.57$. }
\label{fig:fig_8}
\end{figure*}

%lambda=0の場合のvalicdation
\subsubsection{\label{sec:sec412}ARA with $\lambda=0$}
To support the RS ansatz without the static approximation, we consider the ARA with $\lambda=0$.
In this case, the quantum part in Eq. \eqref{E17} disappears. 
The experimental settings are the same as those in Fig. \ref{fig:fig_3}. 
We set $\alpha=0.6$ and $T_0=0$ in Fig. \ref{fig:fig_4a}, and $\alpha=0.62$ and $T_0=0.05$ in Fig. \ref{fig:fig_4b}. 
We consider three initial conditions: $c=0.7$, $0.8$, and 
$0.95$. The initial candidate solutions are generated from Eq. \eqref{E19} given a fixed fraction $c$.  
The error bar is given by the standard deviation.
Each curve represents the RS solutions, and each symbol denotes the numerical results obtained by the Markov-chain Monte Carlo simulations. 
It can be observed that the numerical results are consistent with the RS solutions.
We can see that the deviations between the numerical results and the RS solutions are not the breaking of the RS ansatz to the breaking of the static approximation. 
For $\lambda=0$ with or without noise, the ARA can avoid the first-order phase transition if we prepare for the proper initial conditions. 
In the next Section, we analyze the general cases in detail.

%Analysis  for finite lambda
\subsubsection{\label{sec:sec413}ARA with finite $\lambda$}
We consider the ARA with finite $\lambda$. 
The experimental settings are the same as those in Fig.\ref{fig:fig_1a}. 
Figure \ref{fig:fig_5} presents the phase diagram of the CDMA model in the ARA for $\alpha=0.6$ and $0.5$.
We consider four initial conditions: $c=0.7, 0.8, 0.9$, and $0.95$.
Each curve represents a point of the first-order phase transition. 
We can observe from Figs. \ref{fig:fig_5a} and \ref{fig:fig_5b} that the first-order phase transition can be avoided if the initial state is close to the original signal.
As the information regarding the original signal is increased, the region for avoiding the first-order phase transition is broadened.
In Fig. \ref{fig:fig_5b}, the region in which the first-order phase transition can be avoided is narrower than that in Fig. \ref{fig:fig_5a}. 
For a lower pattern ratio, further information regarding the original signal is initially required to avoid the first-order phase transition. 
We also investigate the stability of the RS solutions and find that the RSB does not happen for finite $\lambda$
\begin{figure*}[t]
\subfigure[\label{fig:fig_9a}]{\includegraphics[width=55mm]{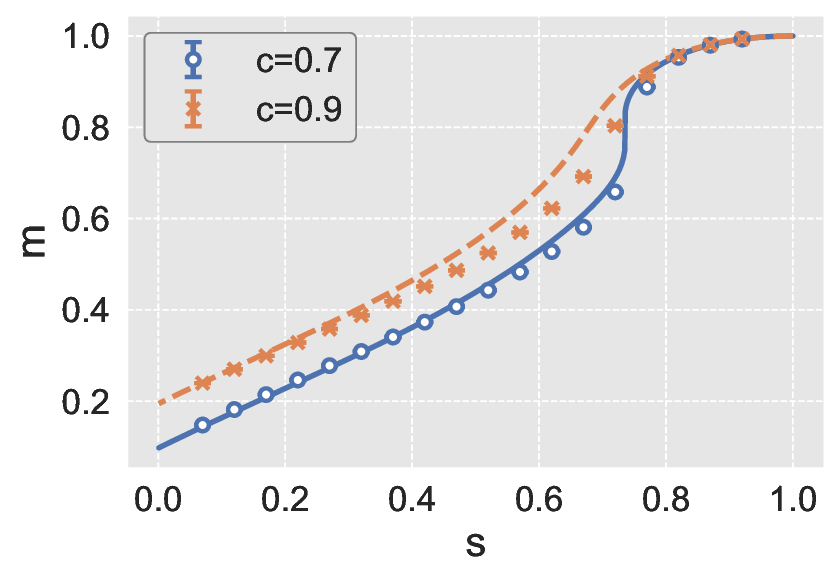}}
\subfigure[\label{fig:fig_9c}]{\includegraphics[width=55mm]{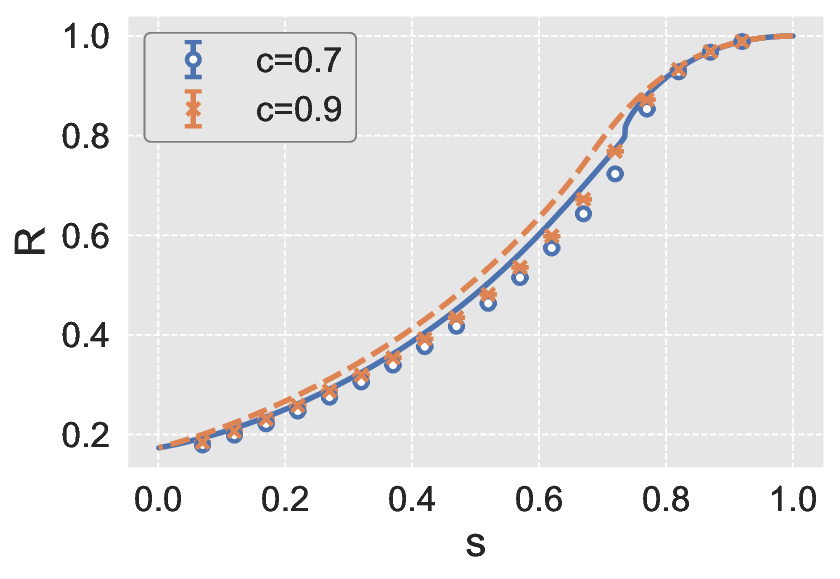}}
\caption{Dependence of order parameters on annealing parameter  with $\lambda=0.8$. 
Both axes are the same as those in Fig. \ref{fig:fig_3}.}
\label{fig:fig_9}
\end{figure*}

%deltamについて
%ARAがなぜうまくいくか
%古典よりいいのはここでいうべき。
To analyze the extent to which the difficulty in obtaining the original signal is mitigated by the ARA, we plot the differences in the magnetization $\Delta m$ between the two local minima at the first-order phase transition in the case of $\alpha=0.6$ and $0.5$ in Fig. \ref{fig:fig_6}.
As discussed in Ref.\cite{ohkuwa_2018}, the rate of quantum tunneling between two local minima in the free energy landscape is related to $\Delta m$.
Figure \ref{fig:fig_6} indicates that $\Delta m$ decreases as $c$ increases. 
For finite $\lambda$, $\Delta m$ is smaller than that of the vanila QA ($\lambda=1$).
Even though the ARA can not eliminate the first-order phase transition,
the two local minima of the free energy become closer than those of the original one. 
The result demonstrates that the ARA enhances the effects of the quantum tunneling for the CDMA model.
In the ARA, we add the bias towards the original signal through the initial Hamiltonian. 
Since the bias removes or softens the free energy barrier, the ARA can avoid or mitigate the first-order phase transition. 

\begin{figure*}[t]
\subfigure[\label{fig:fig_10a}]{\includegraphics[width=55mm]{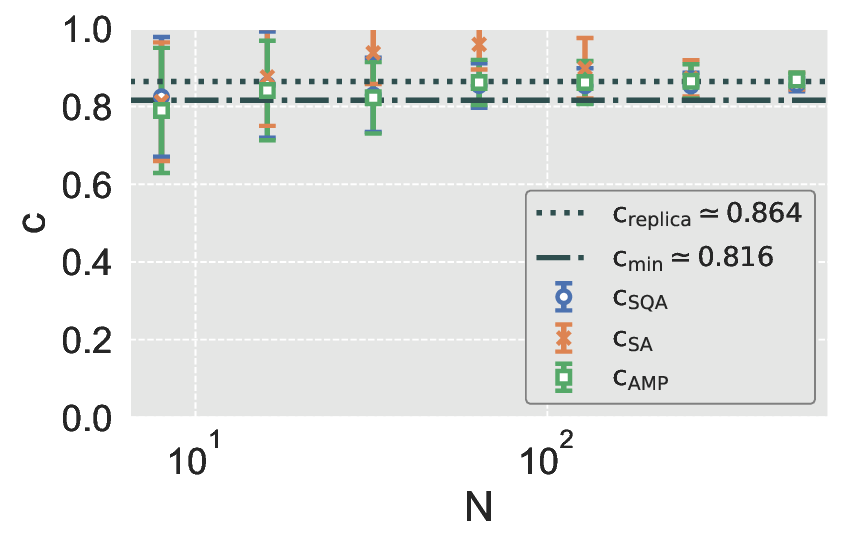}}
\mbox{\raisebox{0mm}{\subfigure[\label{fig:fig_10b}]{\includegraphics[width=55mm]{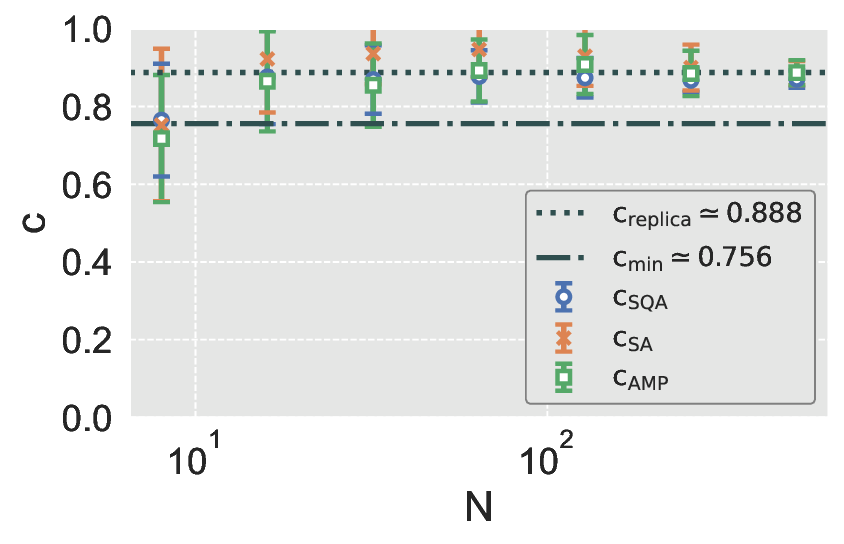}}}}
\caption{The dependence of the fraction of the ground state in the estimated signal obtained by SA, SQA and the AMP algorithm on the system size. 
The experimental settings are as follows: (a)  $\alpha=0.5$ and $T_0=0.0$ and (b) $\alpha=0.57$ and $T_0=0.05$.
}
\label{fig:fig_10}
\end{figure*}

\begin{figure*}[t]
\subfigure[\label{fig:fig_11a}]{\includegraphics[width=55mm]{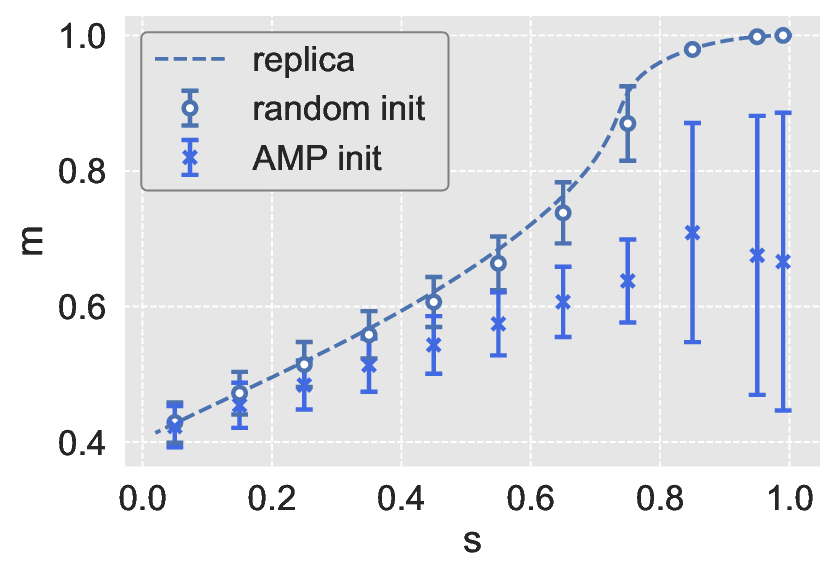}}
\mbox{\raisebox{0mm}{\subfigure[\label{fig:fig_11b}]{\includegraphics[width=55mm]{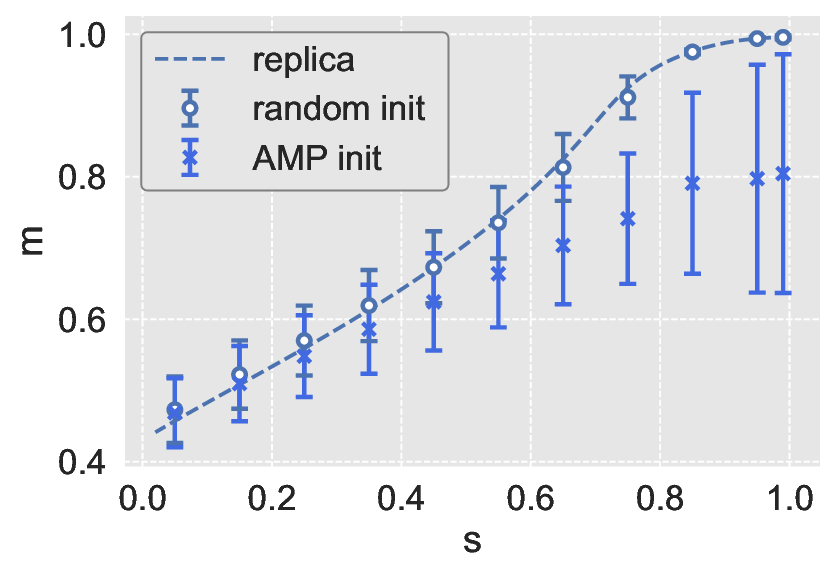}}}}
\caption{Dependence of magnetization on annealing parameter with $\lambda=0.6$ in the ``random init.'' and the ``AMP init.''  settings. 
The dashed curve for each case is obtained by the saddle-point equations.
The experimental settings are as follows: (a) is $\alpha=0.5$ and $T_0=0$, and (b) is $\alpha=0.57$ and $T_0=0.05$. 
Both axes are the same as those in Figs. \ref{fig:fig_3a}.
}
\label{fig:fig_11}
\end{figure*}

%ノイズありの場合
%cases?
We consider the noise effects for the CDMA model in the ARA.
The experimental settings are the same as those illustrated in Fig. \ref{fig:fig_1b}. 
Figure \ref{fig:fig_7} displays the phase diagrams of the CDMA in the ARA for $\alpha=0.62$ and $0.57$.
The qualitative behaviors of the systems are approximately the same as those in the noiseless cases. 
The regions in which the first-order phase transition can be avoided are larger than those of the noiseless cases because the first-order phase transition is weakened owing to the noise effects. 
 Figure \ref{fig:fig_8} presents $\Delta m$ in the case of $\alpha=0.62$ and $0.57$.
 We can see that $\Delta m$ is smaller than in the noiseless cases. 
 In the small noisy cases,the free energy barrier is lower than  in the noiseless cases.
The ARA works better in the small noisy cases than the noiseless cases.

%validation for finite lambda
To validate the RS solutions under the static approximation for finite $\lambda$, we perform quantum Monte Carlo simulations. 
The experimental settings are the same as those in Figs. \ref{fig:fig_3} and \ref{fig:fig_4}. 
We set the RA parameter as $\lambda=0.8$, and the initial conditions as $c=0.7$ and $0.9$.
In Fig.\ref{fig:fig_8}, the order parameters for $c=0.7$ still exhibit a jump.
In the case of $c=0.9$, it can be observed that the first-order phase transition can be avoided.
We can see the deviations between the RS solutions and the numerical results are the same as in Fig.\ref{fig:fig_3}.
In these problem settings, the RSB does not happen from the results of the saddle-point equations. 
Although the numerical results do not entirely match the RS solutions, the qualitative behaviors of the numerical results to avoid the first-order phase transition are similar to those of the RS solutions.

%practical cases
%大まかな地図を示す.
\subsection{\label{sec:sec42}Analysis of the ARA in practical cases}
In  Section.\ref{sec:sec41}, we assume that the initial candidate solution is randomly generated from Eq.\eqref{E19} with a fixed $c$.
Practically, we need to prepare for the initial candidate solution by some algorithms.
At first, we examine whether or not we can prepare for the proper initial condition to avoid the first-order phase transition with commonly used algorithms. 
Next, we evaluate the performance of the ARA with the initial candidate solutions obtained by the algorithms. 
\begin{figure*}[t]
\mbox{\raisebox{0mm}{\subfigure[\label{fig:fig_12a}]{\includegraphics[width=55mm]{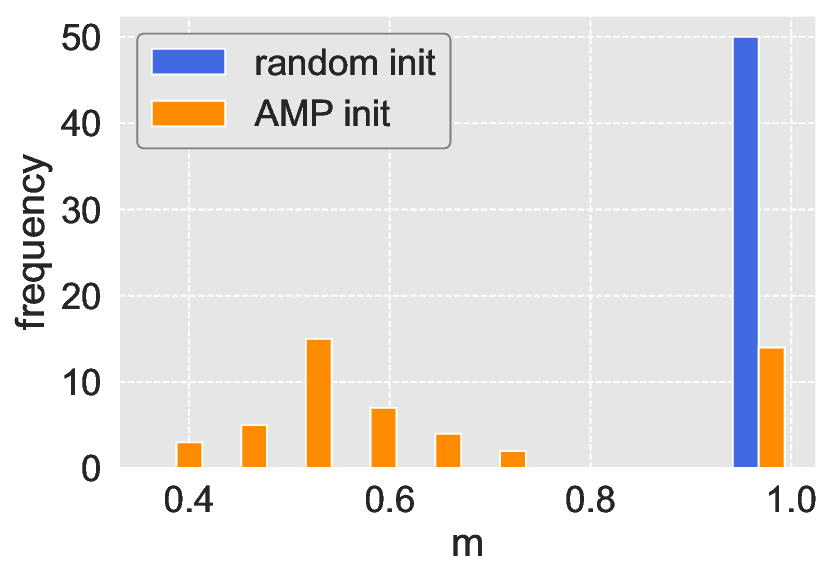}}}}
\subfigure[\label{fig:fig_12b}]{\includegraphics[width=55mm]{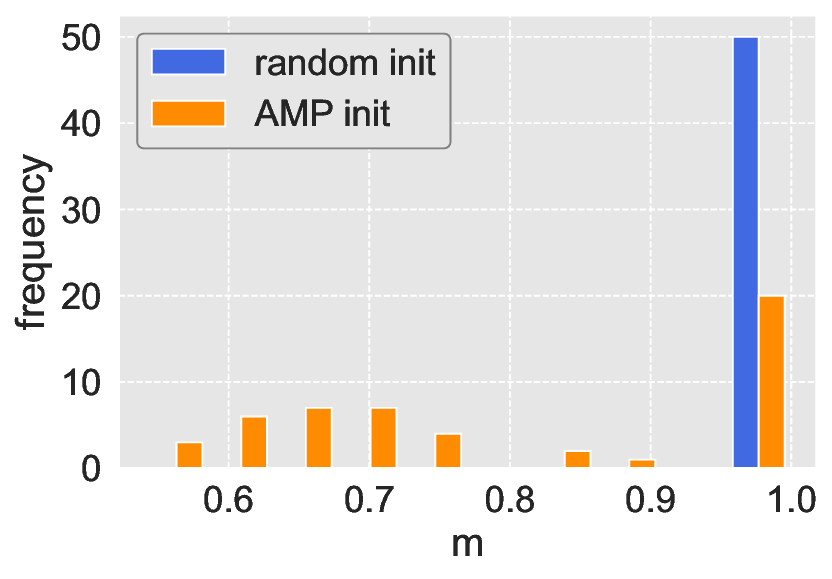}}
\caption{Histogram of magnetization in the ``random init.'' and the ``AMP init.'' settings at $s=0.99$ for $50$ instances in Fig.\ref{fig:fig_11}.
The experimental settings are as follows: (a) is $\alpha=0.5$ and $T_0=0$, and (b) is $\alpha=0.57$ and $T_0=0.05$.}
\label{fig:fig_12}
\end{figure*}

\subsubsection{\label{sec:sec421}How to prepare for the initial candidate solution}
%practical case 
%AMP等で実際に解が求まるのかの実験。
%実験の説明
To prepare for the initial candidate solution, we adopt SA, simulated quantum annealing (SQA), and the approximate message passing (AMP) algorithm \cite{opperbook,Kabashima_2003,Donoho_2009b}.
To perform SA and SQA, we take advantage of OpenJij, an open-source library for heuristic optimization problems in Python \footnote{ \url{https://github.com/OpenJij/OpenJij}}.
The implementation of the AMP algorithm is based on the paper \cite{Kabashima_2003}.
We perform three algorithms for different $50$ instances. For SA and SQA, we carry out $51$ different initial conditions for each instance.
We set the system size as $N=8,16, 32, 64,128,256$ and $512$.
We check the dependence of $c$ on $N$ with these algorithms in Fig.\ref{fig:fig_10}.
We compute $c$ from the relationship $c=(1+\mathcal{M})/2$.
We define the threshold required to avoid the first-order phase transition as $c_{\mathrm{min}}$.
We calculate $c_{\mathrm{min}}$ from the saddle-point equations.
We consider the region where the spinodal curve 2 does not exist, and the critical curve exists, for example $0.45\leq \alpha \leq 0.57$ in Fig.\ref{fig:fig_1}.
In this region, to avoid or mitigate the first-order phase transition is crucial to estimate the original signal efficiently. 

%図の説明
At first, we consider the noise less case : $\alpha=0.5$ and $T_0=0.0$.
In this case, the threshold $c_{\mathrm{min}}$ is nearly equal to $0.816$.
Figure.\ref{fig:fig_10a} shows that the results of the three algorithms almost converge to the RS solutions $c_{\mathrm{replica}}\simeq 0.864$ as we increase $N$. 
These practical algorithms can lead to the candidate solutions exceeding $c_{\mathrm{min}}$.  
Next, we consider the noisy case: $\alpha=0.57$ and $T_0=0.05$. 
The threshold $c_{\mathrm{min}}$ is nearly equal to $0.756$. Figure.\ref{fig:fig_10b} exhibits that these practical algorithms can accomplish $c_{\mathrm{min}}$ in the noisy case.

\subsubsection{\label{sec:sec422}Performance evaluation}
%practical case 
%AMPで得た解の評価。
%実験の説明
We evaluate the performance of the ARA with the initial candidate solution attained by the practical algorithms.
We adopt the AMP algorithm to prepare for the initial candidate solution. 
The experimental settings are the same as those in Fig. \ref{fig:fig_3}. 
%We generate $50$ instances randomly. 
We set the RA parameter as $\lambda=0.6$.
At first, we perform the AMP algorithm for each instance. 
We utilize the final result as the initial candidate solution.
We call this setting  ``AMP init.''.
For the same instance, we randomly generate the initial candidate solution from Eq.\eqref{E19} with a fixed fraction $c$, which is the same as one obtained by the AMP algorithm.
We call this setting  ``random init.''.

%図の説明
We plot the magnetization for two initializations in Fig.\ref{fig:fig_11}.
The error bar is given by the standard deviation.
We consider two cases  $\alpha=0.5$ and $T_0=0$ in Fig. \ref{fig:fig_11a}, and $\alpha=0.57$ and $T_0=0.05$ in Fig. \ref{fig:fig_11b}.
The dashed curves represent the RS solutions with $c_{\mathrm{replica}}$.
In the ``random init.'' setting, the numerical results are consistent with the RS solutions. 
In this setting, the first-order phase transition can be avoided by the ARA with and without noise.
In the ``AMP init.'' setting, the numerical results do not match the RS solutions.
The ARA can not exclude the first-order phase transition.
In Fig.\ref{fig:fig_12},
we plot the histogram of the magnetization at $s=0.99$ used in Fig.\ref{fig:fig_11} to check the existence of the first-order phase transition in detail.
In the ``random init.'' setting,
only one peak exists at $m\simeq 1$.
In the ``AMP init.'' setting, there are two peaks around $m\simeq1$ and $m\neq 1$. 
The ARA can not eliminate the first-order phase transition even though the fraction $c$ obtained by the AMP algorithm exceeds the threshold $c_{\mathrm{min}}$.
For the other practical algorithms, the similar behaviors seem to occur.

%RS解とAMPinitが合わない原因。
The deviations between the numerical results of the ARA in the ``AMP init.'' setting and the RS solutions are due to the assumption of the probability distribution of the initial candidate solution in our replica analysis.
We assume that the probability distribution of the initial candidate solution follows Eq.\eqref{E19}.
Since the initial candidate solution obtained by the AMP algorithm is not generated from Eq.\eqref{E19}, we can not directly apply our analytical results to the ``AMP init.'' setting.

%なぜAMP init.でうまくいくのかrandom initでうまくいくのか
Finally, we consider why the ARA in the ``random init.'' setting can avoid the first-order phase transition and the ARA in the ``AMP init.'' setting can not.
In the ``AMP init.'' setting, the initial candidate solution depends on the original signal, the received signal and the spreading codes.
Therefore, the initial candidate solution is correlated with the received signal and the spreading codes.
Meanwhile, in the ``random init.'' setting, the initial candidate solution depends only on the original signal and does not depend on the received signal and the spreading codes.
Originally, the information about the original signal can be attained through the received signal and the spreading codes.
In the ``random init.'' setting, the initial candidate solution has the information about the original signal not included in the received signal and the spreading codes.
The initial candidate solution increases the effective pattern ratio.
As we increase the pattern ratio, the free energy barrier gets smaller.
Thus, the ARA in the ``random init.'' setting can avoid the first-order phase transition. 
In the ``AMP init.'' setting, the initial candidate solution only has the same information about the original signal obtained through the received signal and the spreading codes.
Since the effective pattern ratio is the same as the original one,
the free energy landscape in the ``AMP init.'' setting is the same as the original one.
Consequently, the ARA in the ``AMP init.'' setting can not eliminate the first-order phase transition even if the candidate solution obtained by the AMP algorithm exceeds $c_{\mathrm{min}}$ which is attained from the saddle-point equations in oracle cases.
To analyze the performance of the ARA in the ``AMP init.'' setting appropriately, we should change the probability distribution of the initial candidate solution in our replica analysis.

\section{\label{sec:sec5}Conclusion}
%全体的なまとめ
We performed a mean-field analysis of the ARA for the CDMA multiuser detection. 
In the CDMA multiuser detection, the first-order phase transition is encountered in the intermediate pattern ratio.
This first-order phase transition degrades the estimation performance. 
To avoid the first-order phase transition, we applied the ARA to the CDMA multiuser detection.

%idealのまとめ
%Firstly, we considered the ARA in oracle cases where the initial candidate solution is randomly generated from Eq.\eqref{E22} with a fixed fraction $c$.
Firstly, we considered the ARA in oracle cases where the initial candidate solution is randomly generated from the original signal with a fixed fraction $c$.
The first-order phase transition can be avoided by the ARA if we prepare for the proper initial condition.
In the ARA, the differences in the magnetization between the two local minima at the first-order phase transition were smaller than those in the vanilla QA. 
The prior information of the original signal avoids or mitigates the first-order phase transition. 
To validate our analysis, we performed quantum Monte Carlo simulations.
The numerical results were consistent with the RS solutions under the static approximation, except for the intermediate values of the annealing parameter.
Although the RS solutions under the static approximation were invalid in these cases,
the results obtained from the RS solutions that the ARA can avoid the first-order phase transition were consistent with the numerical results.
The RS solutions under the static approximation can be useful for understanding the qualitative behaviors of the ARA.

%practicalのまとめ
Next, we considered the ARA in practical cases where we prepare for the initial candidate solution by the practical algorithms.
We considered the three algorithms: SA, SQA, and the AMP algorithms.
The fraction $c$ obtained by these practical algorithms can exceed the threshold $c_{\mathrm{min}}$  to avoid the first-order phase transition.
To evaluate the performance of the ARA with the initial candidate solution attained by the practical algorithms, we performed the quantum Monte Carlo simulations.
In the ``AMP init.'' setting, we prepared for the initial candidate solution with the AMP algorithm.
To compare with the ``AMP init.'' setting, we considered the  ``random init.'' setting where the initial candidate solution was randomly generated from Eq.\eqref{E19} with a fixed fraction $c$, which was the same as one obtained by the AMP algorithm for each instance. 
The ARA in the ``random init.'' setting can utilize the additional information about the original signal not included in the received signal and the spreading codes.
Since the free energy barrier is removed by the additional information about the original signal,
the ARA in the ``random init.'' setting can avoid the first-order phase transition. 
Meanwhile,in the ``AMP init.'' setting, the initial candidate solution was correlated with the received signal and the spreading codes.
The initial candidate solution only had the same information attained through the received signal and the spreading codes.
Because no additional information about the original signal existed, the effective free energy landscape was the same as the original one.
Therefore, the ARA in the ``AMP init.'' setting can not avoid the first-order phase transition.
The ARA in the ``AMP init.'' setting did not match the RS solutions, whereas the ARA in the ``random init.'' setting matched.
The deviations between the numerical result of the ARA in the ``AMP init.'' setting and the RS solutions were due to the assumption of the probability distribution of the initial candidate solution in our replica analysis.

%ここから大切。
In the ``AMP init.'' setting, the initial candidate solution was correlated with the received signal and the spreading codes.
To incorporate the correlation with the received signal and the spreading codes into the initial candidate solution, we need to consider an equilibrium configuration governed by the Gibbs-Boltzmann distribution.
Then, the free energy is constrained by the equilibrium configuration. 
The equilibrium property of the constrained free energy can be analyzed by the Franz-Parisi potential, which is developed to study the metastable state structure for discontinuous mean-field spin glasses \cite{Franz1995,Franz1997,Franz1998,Huang2013,Huang2014}.
In a future study, we will analyze the Franz-Parisi potential for the CDMA to investigate the performance of the ARA in practical cases properly.

Although we can not directly apply our theoretical results to the practical cases, we showed that the ARA in the  ``random init.'' setting can avoid the first-order phase transition for the CDMA multiuser detection.
We exhibited that the probability distribution of the initial candidate solution was crucial in the ARA.
Our results indicated that the effective free energy landscape did not change by the ARA with the initial candidate solution obtained by some practical algorithms.
In practical cases, the ARA did not enhance the estimation performance for the CDMA multiuser detection. The similar behaviors would occur when we apply the ARA to combinatorial optimization problems.
\section*{Acknowledgments}
M.O. was supported by KAKENHI (No. 19H01095 and No. 20H02168), the Next Generation High-
Performance Computing Infrastructures and Applications R $\&$ D Program by MEXT, and MEXT-Quantum Leap Flagship Program Grant Number JPMXS0120352009.
K.T. was supported by JSPS KAKENHI (No. 18H03303). This work was partly supported by JST-CREST (No. JPMJCR1402). 
\appendix
\section{The saddle-point equations and the stability condition of the RS solutions}
\label{appendix_a}
In this appendix, we present the saddle-point equations and the stability condition of the RS solutions.
The extremization of Eq. \eqref{E23} yields the following saddle-point equations: 
\begin{widetext}
\begin{align}
m&=\sum_{a=\pm1}c_a\int DzY_a^{-1}\int Dy\left(\frac{g_a}{u_a}\right)\sinh u_a,\label{A1}\\
q&=\sum_{a=\pm1}c_a\int Dz\left\{Y_a^{-1}\int Dy\left(\frac{g_a}{u_a}\right)\sinh u_a\right\}^2\label{A2}\\
R&=\sum_{a=\pm1}c_a\int DzY_a^{-1}\int Dy\left\{\left(\frac{ (\beta(1-s)\lambda)^2}{u_a^3}\right)\sinh u_a + \left(\frac{g_a}{u_a}\right)^2\cosh u_a\right\},\label{A3}\\
\tilde{m}&=\frac{\alpha \beta s}{1+\beta s(R-q)},\label{A4}\\
\tilde{q}&=\frac{\alpha \beta^2 s^2\left(q-2m+1+\beta_0^{-1}\right)}{(1+\beta s(R-q))^2},\label{A5}\\
2\tilde{R}-\tilde{q}&=\frac{\alpha \beta^2 s^2(R-q)}{1+\beta s(R-q)},\label{A6}\\
Y_{a}&\equiv \int Dy \cosh u_{a},\label{A7}\\
u_a&\equiv \sqrt{g_{a}^2+(\beta(1-s)\lambda)^2}\label{A8}.
\end{align}
The overlap can be written as 
\begin{align}
\mathcal{M}&=\sum_{a=\pm1}c_a\int Dz \mathrm{sgn}\left\{Y_a^{-1}\int Dy\left(\frac{g_a}{u_a}\right)\sinh u_a\right\}\label{A9}.
\end{align}
The fraction of the ground state in the estimated signal is calculated by $c_{\mathrm{repica}}=(1+\mathcal{M})/2$.
Basically we numerical assess the fixed points of the saddle-point equations by iterating the substitution of the tentative solutions.
The fixed points are extremum of Eq. \eqref{E23} and characterize the free energy of the system.

Next, we consider the stability of the RS solutions. In the low-temperature regions, the classical CDMA model exhibits replica symmetry breaking (RSB) \cite{Yoshida2007}.
Two instabilities exist in the RS solutions: 
the local and global instabilities of the RS solutions.
The local stability condition of the RS solutions under the static approximation is expressed as 
\begin{align}
&\frac{\alpha \beta^2 s^2}{(1+\beta s(R-q))^2}\left[\sum_{a=\pm1}c_a\int Dz\left\{\left(Y_a^{-1}\int Dy\left(\frac{g_a}{u_a}\right)\sinh u_a \right)^2-Y_a^{-1}\left(\int Dy \left(\frac{(\beta(1-s)\lambda)^2}{u_a^3}\right)\sinh u_a+\int Dy \left(\frac{g_a}{u_a}\right)^2\cosh u_a \right)\right\}^2\right]<1\label{A11}.
\end{align}
This condition corresponds to the AT condition \cite{Almeida_1978} in the ARA.
This result is consistent with the previous result in Ref. \cite{Yoshida2007} for the classical limit $s=1$ and $\lambda=1$.
We can attain this condition by taking into account the perturbations to the RS solutions \cite{Kabashima_2008,Sakata_2018}. 
The detailed calculations for deriving the AT condition in Eq. \eqref{A11} are presented in \cite{supp1}.
The global instability condition of the RS solutions is related to the negative entropy. The existence of the global instability corresponds to the freezing behavior \cite{Krauth_1989}. 
To detect the freezing behavior, we calculate the RS entropy as follows:
\begin{align}
\mathcal{S}=-\frac{\partial }{\partial T}f_{\mathrm{RS}}&=-\frac{\alpha}{2}\left\{\ln\left(1+\beta s(R-q)\right)\right\}+\frac{R-q}{2}\left(\tilde{m}-\tilde{q}\right)+\tilde{R}R-\frac{1}{2}q\tilde{q}\nonumber\\
&+\sum_{a=\pm1}c_a\int Dz \ln 2Y_a-\beta \left\{\sum_{a=\pm1}c_a\int DzY_a^{-1}\int Dy u_a \sinh u_a \right\}\label{A12}.
\end{align}
In the case of $s=1$ and $\lambda=1$, this result is also consistent with the classical one. 
\end{widetext}
\bibliography{main.bib}

\begin{thebibliography}{81}%
\makeatletter
\providecommand \@ifxundefined [1]{%
 \@ifx{#1\undefined}
}%
\providecommand \@ifnum [1]{%
 \ifnum #1\expandafter \@firstoftwo
 \else \expandafter \@secondoftwo
 \fi
}%
\providecommand \@ifx [1]{%
 \ifx #1\expandafter \@firstoftwo
 \else \expandafter \@secondoftwo
 \fi
}%
\providecommand \natexlab [1]{#1}%
\providecommand \enquote  [1]{``#1''}%
\providecommand \bibnamefont  [1]{#1}%
\providecommand \bibfnamefont [1]{#1}%
\providecommand \citenamefont [1]{#1}%
\providecommand \href@noop [0]{\@secondoftwo}%
\providecommand \href [0]{\begingroup \@sanitize@url \@href}%
\providecommand \@href[1]{\@@startlink{#1}\@@href}%
\providecommand \@@href[1]{\endgroup#1\@@endlink}%
\providecommand \@sanitize@url [0]{\catcode `\\12\catcode `\$12\catcode
  `\&12\catcode `\#12\catcode `\^12\catcode `\_12\catcode `\%12\relax}%
\providecommand \@@startlink[1]{}%
\providecommand \@@endlink[0]{}%
\providecommand \url  [0]{\begingroup\@sanitize@url \@url }%
\providecommand \@url [1]{\endgroup\@href {#1}{\urlprefix }}%
\providecommand \urlprefix  [0]{URL }%
\providecommand \Eprint [0]{\href }%
\providecommand \doibase [0]{http://dx.doi.org/}%
\providecommand \selectlanguage [0]{\@gobble}%
\providecommand \bibinfo  [0]{\@secondoftwo}%
\providecommand \bibfield  [0]{\@secondoftwo}%
\providecommand \translation [1]{[#1]}%
\providecommand \BibitemOpen [0]{}%
\providecommand \bibitemStop [0]{}%
\providecommand \bibitemNoStop [0]{.\EOS\space}%
\providecommand \EOS [0]{\spacefactor3000\relax}%
\providecommand \BibitemShut  [1]{\csname bibitem#1\endcsname}%
\let\auto@bib@innerbib\@empty
%</preamble>
\bibitem [{\citenamefont {Verdu}(1998)}]{Verdu_1998}%
  \BibitemOpen
  \bibfield  {author} {\bibinfo {author} {\bibfnamefont {S.}~\bibnamefont
  {Verdu}},\ }\href@noop {} {\emph {\bibinfo {title} {Multiuser Detection}}},\
  \bibinfo {edition} {1st}\ ed.\ (\bibinfo  {publisher} {Cambridge University
  Press},\ \bibinfo {address} {USA},\ \bibinfo {year} {1998})\BibitemShut
  {NoStop}%
\bibitem [{\citenamefont {Nishimori}(2001)}]{nishimori}%
  \BibitemOpen
  \bibfield  {author} {\bibinfo {author} {\bibfnamefont {H.}~\bibnamefont
  {Nishimori}},\ }\href@noop {} {\emph {\bibinfo {title} {Statistical Physics
  of Spin Glasses and Information Processing: an Introduction}}}\ (\bibinfo
  {publisher} {Oxford University Press},\ \bibinfo {address} {Oxford; New
  York},\ \bibinfo {year} {2001})\BibitemShut {NoStop}%
\bibitem [{\citenamefont {Tanaka}(2001)}]{Tanaka2001}%
  \BibitemOpen
  \bibfield  {author} {\bibinfo {author} {\bibfnamefont {T.}~\bibnamefont
  {Tanaka}},\ }\href {http://stacks.iop.org/0295-5075/54/i=4/a=540} {\bibfield
  {journal} {\bibinfo  {journal} {EPL (Europhysics Letters)}\ }\textbf
  {\bibinfo {volume} {54}},\ \bibinfo {pages} {540} (\bibinfo {year}
  {2001})}\BibitemShut {NoStop}%
\bibitem [{\citenamefont {{Tanaka}}(2002)}]{Tanaka2002}%
  \BibitemOpen
  \bibfield  {author} {\bibinfo {author} {\bibfnamefont {T.}~\bibnamefont
  {{Tanaka}}},\ }\href {\doibase 10.1109/TIT.2002.804053} {\bibfield  {journal}
  {\bibinfo  {journal} {IEEE Transactions on Information Theory}\ }\textbf
  {\bibinfo {volume} {48}},\ \bibinfo {pages} {2888} (\bibinfo {year}
  {2002})}\BibitemShut {NoStop}%
\bibitem [{\citenamefont {Yoshida}\ \emph {et~al.}(2007)\citenamefont
  {Yoshida}, \citenamefont {Uezu}, \citenamefont {Tanaka},\ and\ \citenamefont
  {Okada}}]{Yoshida2007}%
  \BibitemOpen
  \bibfield  {author} {\bibinfo {author} {\bibfnamefont {M.}~\bibnamefont
  {Yoshida}}, \bibinfo {author} {\bibfnamefont {T.}~\bibnamefont {Uezu}},
  \bibinfo {author} {\bibfnamefont {T.}~\bibnamefont {Tanaka}}, \ and\ \bibinfo
  {author} {\bibfnamefont {M.}~\bibnamefont {Okada}},\ }\href {\doibase
  10.1143/JPSJ.76.054003} {\bibfield  {journal} {\bibinfo  {journal} {Journal
  of the Physical Society of Japan}\ }\textbf {\bibinfo {volume} {76}},\
  \bibinfo {pages} {054003} (\bibinfo {year} {2007})}\BibitemShut {NoStop}%
\bibitem [{\citenamefont {{Donoho}}(2006)}]{Donoho_2006}%
  \BibitemOpen
  \bibfield  {author} {\bibinfo {author} {\bibfnamefont {D.~L.}\ \bibnamefont
  {{Donoho}}},\ }\href {\doibase 10.1109/TIT.2006.871582} {\bibfield  {journal}
  {\bibinfo  {journal} {IEEE Transactions on Information Theory}\ }\textbf
  {\bibinfo {volume} {52}},\ \bibinfo {pages} {1289} (\bibinfo {year}
  {2006})}\BibitemShut {NoStop}%
\bibitem [{\citenamefont {Donoho}\ and\ \citenamefont
  {Tanner}(2009)}]{Donoho_2009}%
  \BibitemOpen
  \bibfield  {author} {\bibinfo {author} {\bibfnamefont {D.}~\bibnamefont
  {Donoho}}\ and\ \bibinfo {author} {\bibfnamefont {J.}~\bibnamefont
  {Tanner}},\ }\href {\doibase 10.1098/rsta.2009.0152} {\bibfield  {journal}
  {\bibinfo  {journal} {Philosophical Transactions of the Royal Society A:
  Mathematical, Physical and Engineering Sciences}\ }\textbf {\bibinfo {volume}
  {367}},\ \bibinfo {pages} {4273} (\bibinfo {year} {2009})}\BibitemShut
  {NoStop}%
\bibitem [{\citenamefont {Kabashima}\ \emph {et~al.}(2009)\citenamefont
  {Kabashima}, \citenamefont {Wadayama},\ and\ \citenamefont
  {Tanaka}}]{Kabashima_2009}%
  \BibitemOpen
  \bibfield  {author} {\bibinfo {author} {\bibfnamefont {Y.}~\bibnamefont
  {Kabashima}}, \bibinfo {author} {\bibfnamefont {T.}~\bibnamefont {Wadayama}},
  \ and\ \bibinfo {author} {\bibfnamefont {T.}~\bibnamefont {Tanaka}},\ }\href
  {\doibase 10.1088/1742-5468/2009/09/l09003} {\bibfield  {journal} {\bibinfo
  {journal} {Journal of Statistical Mechanics: Theory and Experiment}\ }\textbf
  {\bibinfo {volume} {2009}},\ \bibinfo {pages} {L09003} (\bibinfo {year}
  {2009})}\BibitemShut {NoStop}%
\bibitem [{\citenamefont {Ganguli}\ and\ \citenamefont
  {Sompolinsky}(2010)}]{Ganguli_2010}%
  \BibitemOpen
  \bibfield  {author} {\bibinfo {author} {\bibfnamefont {S.}~\bibnamefont
  {Ganguli}}\ and\ \bibinfo {author} {\bibfnamefont {H.}~\bibnamefont
  {Sompolinsky}},\ }\href {\doibase 10.1103/PhysRevLett.104.188701} {\bibfield
  {journal} {\bibinfo  {journal} {Phys. Rev. Lett.}\ }\textbf {\bibinfo
  {volume} {104}},\ \bibinfo {pages} {188701} (\bibinfo {year}
  {2010})}\BibitemShut {NoStop}%
\bibitem [{\citenamefont {Inoue}(2001)}]{inoue_2002}%
  \BibitemOpen
  \bibfield  {author} {\bibinfo {author} {\bibfnamefont {J.-i.}\ \bibnamefont
  {Inoue}},\ }\href {\doibase 10.1103/PhysRevE.63.046114} {\bibfield  {journal}
  {\bibinfo  {journal} {Phys. Rev. E}\ }\textbf {\bibinfo {volume} {63}},\
  \bibinfo {pages} {046114} (\bibinfo {year} {2001})}\BibitemShut {NoStop}%
\bibitem [{\citenamefont {Otsubo}\ \emph {et~al.}(2012)\citenamefont {Otsubo},
  \citenamefont {Inoue}, \citenamefont {Nagata},\ and\ \citenamefont
  {Okada}}]{otsubo_2012}%
  \BibitemOpen
  \bibfield  {author} {\bibinfo {author} {\bibfnamefont {Y.}~\bibnamefont
  {Otsubo}}, \bibinfo {author} {\bibfnamefont {J.-i.}\ \bibnamefont {Inoue}},
  \bibinfo {author} {\bibfnamefont {K.}~\bibnamefont {Nagata}}, \ and\ \bibinfo
  {author} {\bibfnamefont {M.}~\bibnamefont {Okada}},\ }\href {\doibase
  10.1103/PhysRevE.86.051138} {\bibfield  {journal} {\bibinfo  {journal} {Phys.
  Rev. E}\ }\textbf {\bibinfo {volume} {86}},\ \bibinfo {pages} {051138}
  (\bibinfo {year} {2012})}\BibitemShut {NoStop}%
\bibitem [{\citenamefont {Otsubo}\ \emph {et~al.}(2014)\citenamefont {Otsubo},
  \citenamefont {Inoue}, \citenamefont {Nagata},\ and\ \citenamefont
  {Okada}}]{otsubo_2014}%
  \BibitemOpen
  \bibfield  {author} {\bibinfo {author} {\bibfnamefont {Y.}~\bibnamefont
  {Otsubo}}, \bibinfo {author} {\bibfnamefont {J.-i.}\ \bibnamefont {Inoue}},
  \bibinfo {author} {\bibfnamefont {K.}~\bibnamefont {Nagata}}, \ and\ \bibinfo
  {author} {\bibfnamefont {M.}~\bibnamefont {Okada}},\ }\href {\doibase
  10.1103/PhysRevE.90.012126} {\bibfield  {journal} {\bibinfo  {journal} {Phys.
  Rev. E}\ }\textbf {\bibinfo {volume} {90}},\ \bibinfo {pages} {012126}
  (\bibinfo {year} {2014})}\BibitemShut {NoStop}%
\bibitem [{\citenamefont {Kadowaki}\ and\ \citenamefont
  {Nishimori}(1998)}]{Kadowaki_1998}%
  \BibitemOpen
  \bibfield  {author} {\bibinfo {author} {\bibfnamefont {T.}~\bibnamefont
  {Kadowaki}}\ and\ \bibinfo {author} {\bibfnamefont {H.}~\bibnamefont
  {Nishimori}},\ }\href {\doibase 10.1103/PhysRevE.58.5355} {\bibfield
  {journal} {\bibinfo  {journal} {Phys. Rev. E}\ }\textbf {\bibinfo {volume}
  {58}},\ \bibinfo {pages} {5355} (\bibinfo {year} {1998})}\BibitemShut
  {NoStop}%
\bibitem [{\citenamefont {Santoro}\ \emph {et~al.}(2002)\citenamefont
  {Santoro}, \citenamefont {Marto{\v n}{\'a}k}, \citenamefont {Tosatti},\ and\
  \citenamefont {Car}}]{Santoro2002}%
  \BibitemOpen
  \bibfield  {author} {\bibinfo {author} {\bibfnamefont {G.~E.}\ \bibnamefont
  {Santoro}}, \bibinfo {author} {\bibfnamefont {R.}~\bibnamefont {Marto{\v
  n}{\'a}k}}, \bibinfo {author} {\bibfnamefont {E.}~\bibnamefont {Tosatti}}, \
  and\ \bibinfo {author} {\bibfnamefont {R.}~\bibnamefont {Car}},\ }\href
  {\doibase 10.1126/science.1068774} {\bibfield  {journal} {\bibinfo  {journal}
  {Science}\ }\textbf {\bibinfo {volume} {295}},\ \bibinfo {pages} {2427}
  (\bibinfo {year} {2002})}\BibitemShut {NoStop}%
\bibitem [{\citenamefont {Santoro}\ and\ \citenamefont
  {Tosatti}(2006)}]{Santoro_2006}%
  \BibitemOpen
  \bibfield  {author} {\bibinfo {author} {\bibfnamefont {G.~E.}\ \bibnamefont
  {Santoro}}\ and\ \bibinfo {author} {\bibfnamefont {E.}~\bibnamefont
  {Tosatti}},\ }\href {\doibase 10.1088/0305-4470/39/36/r01} {\bibfield
  {journal} {\bibinfo  {journal} {Journal of Physics A: Mathematical and
  General}\ }\textbf {\bibinfo {volume} {39}},\ \bibinfo {pages} {R393}
  (\bibinfo {year} {2006})}\BibitemShut {NoStop}%
\bibitem [{\citenamefont {Das}\ and\ \citenamefont
  {Chakrabarti}(2008)}]{Das_2008}%
  \BibitemOpen
  \bibfield  {author} {\bibinfo {author} {\bibfnamefont {A.}~\bibnamefont
  {Das}}\ and\ \bibinfo {author} {\bibfnamefont {B.~K.}\ \bibnamefont
  {Chakrabarti}},\ }\href {\doibase 10.1103/RevModPhys.80.1061} {\bibfield
  {journal} {\bibinfo  {journal} {Rev. Mod. Phys.}\ }\textbf {\bibinfo {volume}
  {80}},\ \bibinfo {pages} {1061} (\bibinfo {year} {2008})}\BibitemShut
  {NoStop}%
\bibitem [{\citenamefont {Morita}\ and\ \citenamefont
  {Nishimori}(2008)}]{Morita_2008}%
  \BibitemOpen
  \bibfield  {author} {\bibinfo {author} {\bibfnamefont {S.}~\bibnamefont
  {Morita}}\ and\ \bibinfo {author} {\bibfnamefont {H.}~\bibnamefont
  {Nishimori}},\ }\href {\doibase 10.1063/1.2995837} {\bibfield  {journal}
  {\bibinfo  {journal} {Journal of Mathematical Physics}\ }\textbf {\bibinfo
  {volume} {49}},\ \bibinfo {pages} {125210} (\bibinfo {year}
  {2008})}\BibitemShut {NoStop}%
\bibitem [{\citenamefont {Somma}\ \emph {et~al.}(2012)\citenamefont {Somma},
  \citenamefont {Nagaj},\ and\ \citenamefont {Kieferov\'a}}]{Somma_2012}%
  \BibitemOpen
  \bibfield  {author} {\bibinfo {author} {\bibfnamefont {R.~D.}\ \bibnamefont
  {Somma}}, \bibinfo {author} {\bibfnamefont {D.}~\bibnamefont {Nagaj}}, \ and\
  \bibinfo {author} {\bibfnamefont {M.}~\bibnamefont {Kieferov\'a}},\ }\href
  {\doibase 10.1103/PhysRevLett.109.050501} {\bibfield  {journal} {\bibinfo
  {journal} {Phys. Rev. Lett.}\ }\textbf {\bibinfo {volume} {109}},\ \bibinfo
  {pages} {050501} (\bibinfo {year} {2012})}\BibitemShut {NoStop}%
\bibitem [{\citenamefont {Farhi}\ \emph {et~al.}(2001)\citenamefont {Farhi},
  \citenamefont {Goldstone}, \citenamefont {Gutmann}, \citenamefont {Lapan},
  \citenamefont {Lundgren},\ and\ \citenamefont {Preda}}]{Farhi_2001}%
  \BibitemOpen
  \bibfield  {author} {\bibinfo {author} {\bibfnamefont {E.}~\bibnamefont
  {Farhi}}, \bibinfo {author} {\bibfnamefont {J.}~\bibnamefont {Goldstone}},
  \bibinfo {author} {\bibfnamefont {S.}~\bibnamefont {Gutmann}}, \bibinfo
  {author} {\bibfnamefont {J.}~\bibnamefont {Lapan}}, \bibinfo {author}
  {\bibfnamefont {A.}~\bibnamefont {Lundgren}}, \ and\ \bibinfo {author}
  {\bibfnamefont {D.}~\bibnamefont {Preda}},\ }\href {\doibase
  10.1126/science.1057726} {\bibfield  {journal} {\bibinfo  {journal}
  {Science}\ }\textbf {\bibinfo {volume} {292}},\ \bibinfo {pages} {472}
  (\bibinfo {year} {2001})}\BibitemShut {NoStop}%
\bibitem [{\citenamefont {Albash}\ and\ \citenamefont
  {Lidar}(2018{\natexlab{a}})}]{Albash_2018}%
  \BibitemOpen
  \bibfield  {author} {\bibinfo {author} {\bibfnamefont {T.}~\bibnamefont
  {Albash}}\ and\ \bibinfo {author} {\bibfnamefont {D.~A.}\ \bibnamefont
  {Lidar}},\ }\href {\doibase 10.1103/RevModPhys.90.015002} {\bibfield
  {journal} {\bibinfo  {journal} {Rev. Mod. Phys.}\ }\textbf {\bibinfo {volume}
  {90}},\ \bibinfo {pages} {015002} (\bibinfo {year}
  {2018}{\natexlab{a}})}\BibitemShut {NoStop}%
\bibitem [{\citenamefont {Denchev}\ \emph {et~al.}(2016)\citenamefont
  {Denchev}, \citenamefont {Boixo}, \citenamefont {Isakov}, \citenamefont
  {Ding}, \citenamefont {Babbush}, \citenamefont {Smelyanskiy}, \citenamefont
  {Martinis},\ and\ \citenamefont {Neven}}]{Denchev_2016}%
  \BibitemOpen
  \bibfield  {author} {\bibinfo {author} {\bibfnamefont {V.~S.}\ \bibnamefont
  {Denchev}}, \bibinfo {author} {\bibfnamefont {S.}~\bibnamefont {Boixo}},
  \bibinfo {author} {\bibfnamefont {S.~V.}\ \bibnamefont {Isakov}}, \bibinfo
  {author} {\bibfnamefont {N.}~\bibnamefont {Ding}}, \bibinfo {author}
  {\bibfnamefont {R.}~\bibnamefont {Babbush}}, \bibinfo {author} {\bibfnamefont
  {V.}~\bibnamefont {Smelyanskiy}}, \bibinfo {author} {\bibfnamefont
  {J.}~\bibnamefont {Martinis}}, \ and\ \bibinfo {author} {\bibfnamefont
  {H.}~\bibnamefont {Neven}},\ }\href {\doibase 10.1103/PhysRevX.6.031015}
  {\bibfield  {journal} {\bibinfo  {journal} {Phys. Rev. X}\ }\textbf {\bibinfo
  {volume} {6}},\ \bibinfo {pages} {031015} (\bibinfo {year}
  {2016})}\BibitemShut {NoStop}%
\bibitem [{\citenamefont {Albash}\ and\ \citenamefont
  {Lidar}(2018{\natexlab{b}})}]{Albash_2018_prx}%
  \BibitemOpen
  \bibfield  {author} {\bibinfo {author} {\bibfnamefont {T.}~\bibnamefont
  {Albash}}\ and\ \bibinfo {author} {\bibfnamefont {D.~A.}\ \bibnamefont
  {Lidar}},\ }\href {\doibase 10.1103/PhysRevX.8.031016} {\bibfield  {journal}
  {\bibinfo  {journal} {Phys. Rev. X}\ }\textbf {\bibinfo {volume} {8}},\
  \bibinfo {pages} {031016} (\bibinfo {year} {2018}{\natexlab{b}})}\BibitemShut
  {NoStop}%
\bibitem [{\citenamefont {Kirkpatrick}\ \emph {et~al.}(1983)\citenamefont
  {Kirkpatrick}, \citenamefont {Gelatt},\ and\ \citenamefont
  {Vecchi}}]{Kirkpatrick_1983}%
  \BibitemOpen
  \bibfield  {author} {\bibinfo {author} {\bibfnamefont {S.}~\bibnamefont
  {Kirkpatrick}}, \bibinfo {author} {\bibfnamefont {C.~D.}\ \bibnamefont
  {Gelatt}}, \ and\ \bibinfo {author} {\bibfnamefont {M.~P.}\ \bibnamefont
  {Vecchi}},\ }\href {\doibase 10.1126/science.220.4598.671} {\bibfield
  {journal} {\bibinfo  {journal} {Science}\ }\textbf {\bibinfo {volume}
  {220}},\ \bibinfo {pages} {671} (\bibinfo {year} {1983})}\BibitemShut
  {NoStop}%
\bibitem [{\citenamefont {Johnson}\ \emph {et~al.}(2010)\citenamefont
  {Johnson}, \citenamefont {Bunyk}, \citenamefont {Maibaum}, \citenamefont
  {Tolkacheva}, \citenamefont {Berkley}, \citenamefont {Chapple}, \citenamefont
  {Harris}, \citenamefont {Johansson}, \citenamefont {Lanting}, \citenamefont
  {Perminov}, \citenamefont {Ladizinsky}, \citenamefont {Oh},\ and\
  \citenamefont {Rose}}]{Dwave2010a}%
  \BibitemOpen
  \bibfield  {author} {\bibinfo {author} {\bibfnamefont {M.~W.}\ \bibnamefont
  {Johnson}}, \bibinfo {author} {\bibfnamefont {P.}~\bibnamefont {Bunyk}},
  \bibinfo {author} {\bibfnamefont {F.}~\bibnamefont {Maibaum}}, \bibinfo
  {author} {\bibfnamefont {E.}~\bibnamefont {Tolkacheva}}, \bibinfo {author}
  {\bibfnamefont {A.~J.}\ \bibnamefont {Berkley}}, \bibinfo {author}
  {\bibfnamefont {E.~M.}\ \bibnamefont {Chapple}}, \bibinfo {author}
  {\bibfnamefont {R.}~\bibnamefont {Harris}}, \bibinfo {author} {\bibfnamefont
  {J.}~\bibnamefont {Johansson}}, \bibinfo {author} {\bibfnamefont
  {T.}~\bibnamefont {Lanting}}, \bibinfo {author} {\bibfnamefont
  {I.}~\bibnamefont {Perminov}}, \bibinfo {author} {\bibfnamefont
  {E.}~\bibnamefont {Ladizinsky}}, \bibinfo {author} {\bibfnamefont
  {T.}~\bibnamefont {Oh}}, \ and\ \bibinfo {author} {\bibfnamefont
  {G.}~\bibnamefont {Rose}},\ }\href@noop {} {\bibfield  {journal} {\bibinfo
  {journal} {Superconductor Science and Technology}\ }\textbf {\bibinfo
  {volume} {23}},\ \bibinfo {pages} {065004} (\bibinfo {year}
  {2010})}\BibitemShut {NoStop}%
\bibitem [{\citenamefont {Berkley}\ \emph {et~al.}(2010)\citenamefont
  {Berkley}, \citenamefont {Johnson}, \citenamefont {Bunyk}, \citenamefont
  {Harris}, \citenamefont {Johansson}, \citenamefont {Lanting}, \citenamefont
  {Ladizinsky}, \citenamefont {Tolkacheva}, \citenamefont {Amin},\ and\
  \citenamefont {Rose}}]{Dwave2010b}%
  \BibitemOpen
  \bibfield  {author} {\bibinfo {author} {\bibfnamefont {A.~J.}\ \bibnamefont
  {Berkley}}, \bibinfo {author} {\bibfnamefont {M.~W.}\ \bibnamefont
  {Johnson}}, \bibinfo {author} {\bibfnamefont {P.}~\bibnamefont {Bunyk}},
  \bibinfo {author} {\bibfnamefont {R.}~\bibnamefont {Harris}}, \bibinfo
  {author} {\bibfnamefont {J.}~\bibnamefont {Johansson}}, \bibinfo {author}
  {\bibfnamefont {T.}~\bibnamefont {Lanting}}, \bibinfo {author} {\bibfnamefont
  {E.}~\bibnamefont {Ladizinsky}}, \bibinfo {author} {\bibfnamefont
  {E.}~\bibnamefont {Tolkacheva}}, \bibinfo {author} {\bibfnamefont {M.~H.~S.}\
  \bibnamefont {Amin}}, \ and\ \bibinfo {author} {\bibfnamefont
  {G.}~\bibnamefont {Rose}},\ }\href@noop {} {\bibfield  {journal} {\bibinfo
  {journal} {Superconductor Science and Technology}\ }\textbf {\bibinfo
  {volume} {23}},\ \bibinfo {pages} {105014} (\bibinfo {year}
  {2010})}\BibitemShut {NoStop}%
\bibitem [{\citenamefont {Harris}\ \emph {et~al.}(2010)\citenamefont {Harris},
  \citenamefont {Johnson}, \citenamefont {Lanting}, \citenamefont {Berkley},
  \citenamefont {Johansson}, \citenamefont {Bunyk}, \citenamefont {Tolkacheva},
  \citenamefont {Ladizinsky}, \citenamefont {Ladizinsky}, \citenamefont {Oh},
  \citenamefont {Cioata}, \citenamefont {Perminov}, \citenamefont {Spear},
  \citenamefont {Enderud}, \citenamefont {Rich}, \citenamefont {Uchaikin},
  \citenamefont {Thom}, \citenamefont {Chapple}, \citenamefont {Wang},
  \citenamefont {Wilson}, \citenamefont {Amin}, \citenamefont {Dickson},
  \citenamefont {Karimi}, \citenamefont {Macready}, \citenamefont {Truncik},\
  and\ \citenamefont {Rose}}]{Dwave2010c}%
  \BibitemOpen
  \bibfield  {author} {\bibinfo {author} {\bibfnamefont {R.}~\bibnamefont
  {Harris}}, \bibinfo {author} {\bibfnamefont {M.~W.}\ \bibnamefont {Johnson}},
  \bibinfo {author} {\bibfnamefont {T.}~\bibnamefont {Lanting}}, \bibinfo
  {author} {\bibfnamefont {A.~J.}\ \bibnamefont {Berkley}}, \bibinfo {author}
  {\bibfnamefont {J.}~\bibnamefont {Johansson}}, \bibinfo {author}
  {\bibfnamefont {P.}~\bibnamefont {Bunyk}}, \bibinfo {author} {\bibfnamefont
  {E.}~\bibnamefont {Tolkacheva}}, \bibinfo {author} {\bibfnamefont
  {E.}~\bibnamefont {Ladizinsky}}, \bibinfo {author} {\bibfnamefont
  {N.}~\bibnamefont {Ladizinsky}}, \bibinfo {author} {\bibfnamefont
  {T.}~\bibnamefont {Oh}}, \bibinfo {author} {\bibfnamefont {F.}~\bibnamefont
  {Cioata}}, \bibinfo {author} {\bibfnamefont {I.}~\bibnamefont {Perminov}},
  \bibinfo {author} {\bibfnamefont {P.}~\bibnamefont {Spear}}, \bibinfo
  {author} {\bibfnamefont {C.}~\bibnamefont {Enderud}}, \bibinfo {author}
  {\bibfnamefont {C.}~\bibnamefont {Rich}}, \bibinfo {author} {\bibfnamefont
  {S.}~\bibnamefont {Uchaikin}}, \bibinfo {author} {\bibfnamefont {M.~C.}\
  \bibnamefont {Thom}}, \bibinfo {author} {\bibfnamefont {E.~M.}\ \bibnamefont
  {Chapple}}, \bibinfo {author} {\bibfnamefont {J.}~\bibnamefont {Wang}},
  \bibinfo {author} {\bibfnamefont {B.}~\bibnamefont {Wilson}}, \bibinfo
  {author} {\bibfnamefont {M.~H.~S.}\ \bibnamefont {Amin}}, \bibinfo {author}
  {\bibfnamefont {N.}~\bibnamefont {Dickson}}, \bibinfo {author} {\bibfnamefont
  {K.}~\bibnamefont {Karimi}}, \bibinfo {author} {\bibfnamefont
  {B.}~\bibnamefont {Macready}}, \bibinfo {author} {\bibfnamefont {C.~J.~S.}\
  \bibnamefont {Truncik}}, \ and\ \bibinfo {author} {\bibfnamefont
  {G.}~\bibnamefont {Rose}},\ }\href {\doibase 10.1103/PhysRevB.82.024511}
  {\bibfield  {journal} {\bibinfo  {journal} {Phys. Rev. B}\ }\textbf {\bibinfo
  {volume} {82}},\ \bibinfo {pages} {024511} (\bibinfo {year}
  {2010})}\BibitemShut {NoStop}%
\bibitem [{\citenamefont {Boixo}\ \emph {et~al.}(2014)\citenamefont {Boixo},
  \citenamefont {R{\o}nnow}, \citenamefont {Isakov}, \citenamefont {Wang},
  \citenamefont {Wecker}, \citenamefont {Lidar}, \citenamefont {Martinis},\
  and\ \citenamefont {Troyer}}]{Dwave2014a}%
  \BibitemOpen
  \bibfield  {author} {\bibinfo {author} {\bibfnamefont {S.}~\bibnamefont
  {Boixo}}, \bibinfo {author} {\bibfnamefont {T.~F.}\ \bibnamefont
  {R{\o}nnow}}, \bibinfo {author} {\bibfnamefont {S.~V.}\ \bibnamefont
  {Isakov}}, \bibinfo {author} {\bibfnamefont {Z.}~\bibnamefont {Wang}},
  \bibinfo {author} {\bibfnamefont {D.}~\bibnamefont {Wecker}}, \bibinfo
  {author} {\bibfnamefont {D.~A.}\ \bibnamefont {Lidar}}, \bibinfo {author}
  {\bibfnamefont {J.~M.}\ \bibnamefont {Martinis}}, \ and\ \bibinfo {author}
  {\bibfnamefont {M.}~\bibnamefont {Troyer}},\ }\href
  {http://dx.doi.org/10.1038/nphys2900} {\bibfield  {journal} {\bibinfo
  {journal} {Nature Physics}\ }\textbf {\bibinfo {volume} {10}},\ \bibinfo
  {pages} {218 EP } (\bibinfo {year} {2014})}\BibitemShut {NoStop}%
\bibitem [{\citenamefont {Katzgraber}\ \emph {et~al.}(2014)\citenamefont
  {Katzgraber}, \citenamefont {Hamze},\ and\ \citenamefont
  {Andrist}}]{Dwave2014b}%
  \BibitemOpen
  \bibfield  {author} {\bibinfo {author} {\bibfnamefont {H.~G.}\ \bibnamefont
  {Katzgraber}}, \bibinfo {author} {\bibfnamefont {F.}~\bibnamefont {Hamze}}, \
  and\ \bibinfo {author} {\bibfnamefont {R.~S.}\ \bibnamefont {Andrist}},\
  }\href {\doibase 10.1103/PhysRevX.4.021008} {\bibfield  {journal} {\bibinfo
  {journal} {Phys. Rev. X}\ }\textbf {\bibinfo {volume} {4}},\ \bibinfo {pages}
  {021008} (\bibinfo {year} {2014})}\BibitemShut {NoStop}%
\bibitem [{\citenamefont {Rosenberg}\ \emph {et~al.}(2016)\citenamefont
  {Rosenberg}, \citenamefont {Haghnegahdar}, \citenamefont {Goddard},
  \citenamefont {Carr}, \citenamefont {Wu},\ and\ \citenamefont
  {de~Prado}}]{Rosenberg_2016}%
  \BibitemOpen
  \bibfield  {author} {\bibinfo {author} {\bibfnamefont {G.}~\bibnamefont
  {Rosenberg}}, \bibinfo {author} {\bibfnamefont {P.}~\bibnamefont
  {Haghnegahdar}}, \bibinfo {author} {\bibfnamefont {P.}~\bibnamefont
  {Goddard}}, \bibinfo {author} {\bibfnamefont {P.}~\bibnamefont {Carr}},
  \bibinfo {author} {\bibfnamefont {K.}~\bibnamefont {Wu}}, \ and\ \bibinfo
  {author} {\bibfnamefont {M.~L.}\ \bibnamefont {de~Prado}},\ }\href {\doibase
  10.1109/JSTSP.2016.2574703} {\bibfield  {journal} {\bibinfo  {journal} {IEEE
  Journal of Selected Topics in Signal Processing}\ }\textbf {\bibinfo {volume}
  {10}},\ \bibinfo {pages} {1053} (\bibinfo {year} {2016})}\BibitemShut
  {NoStop}%
\bibitem [{\citenamefont {Venturelli}\ and\ \citenamefont
  {Kondratyev}(2019)}]{Venturelli_2019}%
  \BibitemOpen
  \bibfield  {author} {\bibinfo {author} {\bibfnamefont {D.}~\bibnamefont
  {Venturelli}}\ and\ \bibinfo {author} {\bibfnamefont {A.}~\bibnamefont
  {Kondratyev}},\ }\href {\doibase 10.1007/s42484-019-00001-w} {\bibfield
  {journal} {\bibinfo  {journal} {Quantum Machine Intelligence}\ }\textbf
  {\bibinfo {volume} {1}},\ \bibinfo {pages} {17} (\bibinfo {year}
  {2019})}\BibitemShut {NoStop}%
\bibitem [{\citenamefont {Neukart}\ \emph {et~al.}(2017)\citenamefont
  {Neukart}, \citenamefont {Compostella}, \citenamefont {Seidel}, \citenamefont
  {von Dollen}, \citenamefont {Yarkoni},\ and\ \citenamefont
  {Parney}}]{Neukart2017}%
  \BibitemOpen
  \bibfield  {author} {\bibinfo {author} {\bibfnamefont {F.}~\bibnamefont
  {Neukart}}, \bibinfo {author} {\bibfnamefont {G.}~\bibnamefont
  {Compostella}}, \bibinfo {author} {\bibfnamefont {C.}~\bibnamefont {Seidel}},
  \bibinfo {author} {\bibfnamefont {D.}~\bibnamefont {von Dollen}}, \bibinfo
  {author} {\bibfnamefont {S.}~\bibnamefont {Yarkoni}}, \ and\ \bibinfo
  {author} {\bibfnamefont {B.}~\bibnamefont {Parney}},\ }\href@noop {}
  {\bibfield  {journal} {\bibinfo  {journal} {Frontiers in ICT}\ }\textbf
  {\bibinfo {volume} {4}},\ \bibinfo {pages} {29} (\bibinfo {year}
  {2017})}\BibitemShut {NoStop}%
\bibitem [{\citenamefont {Nishimura}\ \emph {et~al.}(2019)\citenamefont
  {Nishimura}, \citenamefont {Tanahashi}, \citenamefont {Suganuma},
  \citenamefont {Miyama},\ and\ \citenamefont {Ohzeki}}]{Nishimura_2019}%
  \BibitemOpen
  \bibfield  {author} {\bibinfo {author} {\bibfnamefont {N.}~\bibnamefont
  {Nishimura}}, \bibinfo {author} {\bibfnamefont {K.}~\bibnamefont
  {Tanahashi}}, \bibinfo {author} {\bibfnamefont {K.}~\bibnamefont {Suganuma}},
  \bibinfo {author} {\bibfnamefont {M.~J.}\ \bibnamefont {Miyama}}, \ and\
  \bibinfo {author} {\bibfnamefont {M.}~\bibnamefont {Ohzeki}},\ }\href
  {\doibase 10.3389/fcomp.2019.00002} {\bibfield  {journal} {\bibinfo
  {journal} {Frontiers in Computer Science}\ }\textbf {\bibinfo {volume} {1}},\
  \bibinfo {pages} {2} (\bibinfo {year} {2019})}\BibitemShut {NoStop}%
\bibitem [{\citenamefont {Ohzeki}\ \emph {et~al.}(2019)\citenamefont {Ohzeki},
  \citenamefont {Miki}, \citenamefont {Miyama},\ and\ \citenamefont
  {Terabe}}]{Ohzeki_2019}%
  \BibitemOpen
  \bibfield  {author} {\bibinfo {author} {\bibfnamefont {M.}~\bibnamefont
  {Ohzeki}}, \bibinfo {author} {\bibfnamefont {A.}~\bibnamefont {Miki}},
  \bibinfo {author} {\bibfnamefont {M.~J.}\ \bibnamefont {Miyama}}, \ and\
  \bibinfo {author} {\bibfnamefont {M.}~\bibnamefont {Terabe}},\ }\href
  {\doibase 10.3389/fcomp.2019.00009} {\bibfield  {journal} {\bibinfo
  {journal} {Frontiers in Computer Science}\ }\textbf {\bibinfo {volume} {1}},\
  \bibinfo {pages} {9} (\bibinfo {year} {2019})}\BibitemShut {NoStop}%
\bibitem [{\citenamefont {{Crawford}}\ \emph {et~al.}(2016)\citenamefont
  {{Crawford}}, \citenamefont {{Levit}}, \citenamefont {{Ghadermarzy}},
  \citenamefont {{Oberoi}},\ and\ \citenamefont {{Ronagh}}}]{Crawford2016}%
  \BibitemOpen
  \bibfield  {author} {\bibinfo {author} {\bibfnamefont {D.}~\bibnamefont
  {{Crawford}}}, \bibinfo {author} {\bibfnamefont {A.}~\bibnamefont {{Levit}}},
  \bibinfo {author} {\bibfnamefont {N.}~\bibnamefont {{Ghadermarzy}}}, \bibinfo
  {author} {\bibfnamefont {J.~S.}\ \bibnamefont {{Oberoi}}}, \ and\ \bibinfo
  {author} {\bibfnamefont {P.}~\bibnamefont {{Ronagh}}},\ }\href@noop {}
  {\bibfield  {journal} {\bibinfo  {journal} {ArXiv e-prints}\ } (\bibinfo
  {year} {2016})},\ \Eprint {http://arxiv.org/abs/1612.05695} {arXiv:1612.05695
  [quant-ph]} \BibitemShut {NoStop}%
\bibitem [{\citenamefont {Benedetti}\ \emph {et~al.}(2016)\citenamefont
  {Benedetti}, \citenamefont {Realpe-G\'omez}, \citenamefont {Biswas},\ and\
  \citenamefont {Perdomo-Ortiz}}]{Benedetti_2016}%
  \BibitemOpen
  \bibfield  {author} {\bibinfo {author} {\bibfnamefont {M.}~\bibnamefont
  {Benedetti}}, \bibinfo {author} {\bibfnamefont {J.}~\bibnamefont
  {Realpe-G\'omez}}, \bibinfo {author} {\bibfnamefont {R.}~\bibnamefont
  {Biswas}}, \ and\ \bibinfo {author} {\bibfnamefont {A.}~\bibnamefont
  {Perdomo-Ortiz}},\ }\href {\doibase 10.1103/PhysRevA.94.022308} {\bibfield
  {journal} {\bibinfo  {journal} {Phys. Rev. A}\ }\textbf {\bibinfo {volume}
  {94}},\ \bibinfo {pages} {022308} (\bibinfo {year} {2016})}\BibitemShut
  {NoStop}%
\bibitem [{\citenamefont {Neukart}\ \emph {et~al.}(2018)\citenamefont
  {Neukart}, \citenamefont {Von~Dollen}, \citenamefont {Seidel},\ and\
  \citenamefont {Compostella}}]{Neukart_2018}%
  \BibitemOpen
  \bibfield  {author} {\bibinfo {author} {\bibfnamefont {F.}~\bibnamefont
  {Neukart}}, \bibinfo {author} {\bibfnamefont {D.}~\bibnamefont {Von~Dollen}},
  \bibinfo {author} {\bibfnamefont {C.}~\bibnamefont {Seidel}}, \ and\ \bibinfo
  {author} {\bibfnamefont {G.}~\bibnamefont {Compostella}},\ }\href {\doibase
  10.3389/fphy.2017.00071} {\bibfield  {journal} {\bibinfo  {journal}
  {Frontiers in Physics}\ }\textbf {\bibinfo {volume} {5}},\ \bibinfo {pages}
  {71} (\bibinfo {year} {2018})}\BibitemShut {NoStop}%
\bibitem [{\citenamefont {Amin}\ \emph {et~al.}(2018)\citenamefont {Amin},
  \citenamefont {Andriyash}, \citenamefont {Rolfe}, \citenamefont
  {Kulchytskyy},\ and\ \citenamefont {Melko}}]{Amin_2018}%
  \BibitemOpen
  \bibfield  {author} {\bibinfo {author} {\bibfnamefont {M.~H.}\ \bibnamefont
  {Amin}}, \bibinfo {author} {\bibfnamefont {E.}~\bibnamefont {Andriyash}},
  \bibinfo {author} {\bibfnamefont {J.}~\bibnamefont {Rolfe}}, \bibinfo
  {author} {\bibfnamefont {B.}~\bibnamefont {Kulchytskyy}}, \ and\ \bibinfo
  {author} {\bibfnamefont {R.}~\bibnamefont {Melko}},\ }\href {\doibase
  10.1103/PhysRevX.8.021050} {\bibfield  {journal} {\bibinfo  {journal} {Phys.
  Rev. X}\ }\textbf {\bibinfo {volume} {8}},\ \bibinfo {pages} {021050}
  (\bibinfo {year} {2018})}\BibitemShut {NoStop}%
\bibitem [{\citenamefont {Khoshaman}\ \emph {et~al.}(2018)\citenamefont
  {Khoshaman}, \citenamefont {Vinci}, \citenamefont {Denis}, \citenamefont
  {Andriyash}, \citenamefont {Sadeghi},\ and\ \citenamefont
  {Amin}}]{Khoshaman_2018}%
  \BibitemOpen
  \bibfield  {author} {\bibinfo {author} {\bibfnamefont {A.}~\bibnamefont
  {Khoshaman}}, \bibinfo {author} {\bibfnamefont {W.}~\bibnamefont {Vinci}},
  \bibinfo {author} {\bibfnamefont {B.}~\bibnamefont {Denis}}, \bibinfo
  {author} {\bibfnamefont {E.}~\bibnamefont {Andriyash}}, \bibinfo {author}
  {\bibfnamefont {H.}~\bibnamefont {Sadeghi}}, \ and\ \bibinfo {author}
  {\bibfnamefont {M.~H.}\ \bibnamefont {Amin}},\ }\href {\doibase
  10.1088/2058-9565/aada1f} {\bibfield  {journal} {\bibinfo  {journal} {Quantum
  Science and Technology}\ }\textbf {\bibinfo {volume} {4}},\ \bibinfo {pages}
  {014001} (\bibinfo {year} {2018})}\BibitemShut {NoStop}%
\bibitem [{\citenamefont {King}\ \emph {et~al.}(2018)\citenamefont {King},
  \citenamefont {Carrasquilla}, \citenamefont {Raymond}, \citenamefont
  {Ozfidan}, \citenamefont {Andriyash}, \citenamefont {Berkley}, \citenamefont
  {Reis}, \citenamefont {Lanting}, \citenamefont {Harris}, \citenamefont
  {Altomare}, \citenamefont {Boothby}, \citenamefont {Bunyk}, \citenamefont
  {Enderud}, \citenamefont {Fr{\'e}chette}, \citenamefont {Hoskinson},
  \citenamefont {Ladizinsky}, \citenamefont {Oh}, \citenamefont
  {Poulin-Lamarre}, \citenamefont {Rich}, \citenamefont {Sato}, \citenamefont
  {Smirnov}, \citenamefont {Swenson}, \citenamefont {Volkmann}, \citenamefont
  {Whittaker}, \citenamefont {Yao}, \citenamefont {Ladizinsky}, \citenamefont
  {Johnson}, \citenamefont {Hilton},\ and\ \citenamefont {Amin}}]{King_2018}%
  \BibitemOpen
  \bibfield  {author} {\bibinfo {author} {\bibfnamefont {A.~D.}\ \bibnamefont
  {King}}, \bibinfo {author} {\bibfnamefont {J.}~\bibnamefont {Carrasquilla}},
  \bibinfo {author} {\bibfnamefont {J.}~\bibnamefont {Raymond}}, \bibinfo
  {author} {\bibfnamefont {I.}~\bibnamefont {Ozfidan}}, \bibinfo {author}
  {\bibfnamefont {E.}~\bibnamefont {Andriyash}}, \bibinfo {author}
  {\bibfnamefont {A.}~\bibnamefont {Berkley}}, \bibinfo {author} {\bibfnamefont
  {M.}~\bibnamefont {Reis}}, \bibinfo {author} {\bibfnamefont {T.}~\bibnamefont
  {Lanting}}, \bibinfo {author} {\bibfnamefont {R.}~\bibnamefont {Harris}},
  \bibinfo {author} {\bibfnamefont {F.}~\bibnamefont {Altomare}}, \bibinfo
  {author} {\bibfnamefont {K.}~\bibnamefont {Boothby}}, \bibinfo {author}
  {\bibfnamefont {P.~I.}\ \bibnamefont {Bunyk}}, \bibinfo {author}
  {\bibfnamefont {C.}~\bibnamefont {Enderud}}, \bibinfo {author} {\bibfnamefont
  {A.}~\bibnamefont {Fr{\'e}chette}}, \bibinfo {author} {\bibfnamefont
  {E.}~\bibnamefont {Hoskinson}}, \bibinfo {author} {\bibfnamefont
  {N.}~\bibnamefont {Ladizinsky}}, \bibinfo {author} {\bibfnamefont
  {T.}~\bibnamefont {Oh}}, \bibinfo {author} {\bibfnamefont {G.}~\bibnamefont
  {Poulin-Lamarre}}, \bibinfo {author} {\bibfnamefont {C.}~\bibnamefont
  {Rich}}, \bibinfo {author} {\bibfnamefont {Y.}~\bibnamefont {Sato}}, \bibinfo
  {author} {\bibfnamefont {A.~Y.}\ \bibnamefont {Smirnov}}, \bibinfo {author}
  {\bibfnamefont {L.~J.}\ \bibnamefont {Swenson}}, \bibinfo {author}
  {\bibfnamefont {M.~H.}\ \bibnamefont {Volkmann}}, \bibinfo {author}
  {\bibfnamefont {J.}~\bibnamefont {Whittaker}}, \bibinfo {author}
  {\bibfnamefont {J.}~\bibnamefont {Yao}}, \bibinfo {author} {\bibfnamefont
  {E.}~\bibnamefont {Ladizinsky}}, \bibinfo {author} {\bibfnamefont {M.~W.}\
  \bibnamefont {Johnson}}, \bibinfo {author} {\bibfnamefont {J.}~\bibnamefont
  {Hilton}}, \ and\ \bibinfo {author} {\bibfnamefont {M.~H.}\ \bibnamefont
  {Amin}},\ }\href {\doibase 10.1038/s41586-018-0410-x} {\bibfield  {journal}
  {\bibinfo  {journal} {Nature}\ }\textbf {\bibinfo {volume} {560}},\ \bibinfo
  {pages} {456} (\bibinfo {year} {2018})}\BibitemShut {NoStop}%
\bibitem [{\citenamefont {Harris}\ \emph {et~al.}(2018)\citenamefont {Harris},
  \citenamefont {Sato}, \citenamefont {Berkley}, \citenamefont {Reis},
  \citenamefont {Altomare}, \citenamefont {Amin}, \citenamefont {Boothby},
  \citenamefont {Bunyk}, \citenamefont {Deng}, \citenamefont {Enderud},
  \citenamefont {Huang}, \citenamefont {Hoskinson}, \citenamefont {Johnson},
  \citenamefont {Ladizinsky}, \citenamefont {Ladizinsky}, \citenamefont
  {Lanting}, \citenamefont {Li}, \citenamefont {Medina}, \citenamefont
  {Molavi}, \citenamefont {Neufeld}, \citenamefont {Oh}, \citenamefont
  {Pavlov}, \citenamefont {Perminov}, \citenamefont {Poulin-Lamarre},
  \citenamefont {Rich}, \citenamefont {Smirnov}, \citenamefont {Swenson},
  \citenamefont {Tsai}, \citenamefont {Volkmann}, \citenamefont {Whittaker},\
  and\ \citenamefont {Yao}}]{Harris_2018}%
  \BibitemOpen
  \bibfield  {author} {\bibinfo {author} {\bibfnamefont {R.}~\bibnamefont
  {Harris}}, \bibinfo {author} {\bibfnamefont {Y.}~\bibnamefont {Sato}},
  \bibinfo {author} {\bibfnamefont {A.~J.}\ \bibnamefont {Berkley}}, \bibinfo
  {author} {\bibfnamefont {M.}~\bibnamefont {Reis}}, \bibinfo {author}
  {\bibfnamefont {F.}~\bibnamefont {Altomare}}, \bibinfo {author}
  {\bibfnamefont {M.~H.}\ \bibnamefont {Amin}}, \bibinfo {author}
  {\bibfnamefont {K.}~\bibnamefont {Boothby}}, \bibinfo {author} {\bibfnamefont
  {P.}~\bibnamefont {Bunyk}}, \bibinfo {author} {\bibfnamefont
  {C.}~\bibnamefont {Deng}}, \bibinfo {author} {\bibfnamefont {C.}~\bibnamefont
  {Enderud}}, \bibinfo {author} {\bibfnamefont {S.}~\bibnamefont {Huang}},
  \bibinfo {author} {\bibfnamefont {E.}~\bibnamefont {Hoskinson}}, \bibinfo
  {author} {\bibfnamefont {M.~W.}\ \bibnamefont {Johnson}}, \bibinfo {author}
  {\bibfnamefont {E.}~\bibnamefont {Ladizinsky}}, \bibinfo {author}
  {\bibfnamefont {N.}~\bibnamefont {Ladizinsky}}, \bibinfo {author}
  {\bibfnamefont {T.}~\bibnamefont {Lanting}}, \bibinfo {author} {\bibfnamefont
  {R.}~\bibnamefont {Li}}, \bibinfo {author} {\bibfnamefont {T.}~\bibnamefont
  {Medina}}, \bibinfo {author} {\bibfnamefont {R.}~\bibnamefont {Molavi}},
  \bibinfo {author} {\bibfnamefont {R.}~\bibnamefont {Neufeld}}, \bibinfo
  {author} {\bibfnamefont {T.}~\bibnamefont {Oh}}, \bibinfo {author}
  {\bibfnamefont {I.}~\bibnamefont {Pavlov}}, \bibinfo {author} {\bibfnamefont
  {I.}~\bibnamefont {Perminov}}, \bibinfo {author} {\bibfnamefont
  {G.}~\bibnamefont {Poulin-Lamarre}}, \bibinfo {author} {\bibfnamefont
  {C.}~\bibnamefont {Rich}}, \bibinfo {author} {\bibfnamefont {A.}~\bibnamefont
  {Smirnov}}, \bibinfo {author} {\bibfnamefont {L.}~\bibnamefont {Swenson}},
  \bibinfo {author} {\bibfnamefont {N.}~\bibnamefont {Tsai}}, \bibinfo {author}
  {\bibfnamefont {M.}~\bibnamefont {Volkmann}}, \bibinfo {author}
  {\bibfnamefont {J.}~\bibnamefont {Whittaker}}, \ and\ \bibinfo {author}
  {\bibfnamefont {J.}~\bibnamefont {Yao}},\ }\href {\doibase
  10.1126/science.aat2025} {\bibfield  {journal} {\bibinfo  {journal}
  {Science}\ }\textbf {\bibinfo {volume} {361}},\ \bibinfo {pages} {162}
  (\bibinfo {year} {2018})}\BibitemShut {NoStop}%
\bibitem [{\citenamefont {Weinberg}\ \emph {et~al.}(2020)\citenamefont
  {Weinberg}, \citenamefont {Tylutki}, \citenamefont {R\"onkk\"o},
  \citenamefont {Westerholm}, \citenamefont {\AA{}str\"om}, \citenamefont
  {Manninen}, \citenamefont {T\"orm\"a},\ and\ \citenamefont
  {Sandvik}}]{Weinberg_2020}%
  \BibitemOpen
  \bibfield  {author} {\bibinfo {author} {\bibfnamefont {P.}~\bibnamefont
  {Weinberg}}, \bibinfo {author} {\bibfnamefont {M.}~\bibnamefont {Tylutki}},
  \bibinfo {author} {\bibfnamefont {J.~M.}\ \bibnamefont {R\"onkk\"o}},
  \bibinfo {author} {\bibfnamefont {J.}~\bibnamefont {Westerholm}}, \bibinfo
  {author} {\bibfnamefont {J.~A.}\ \bibnamefont {\AA{}str\"om}}, \bibinfo
  {author} {\bibfnamefont {P.}~\bibnamefont {Manninen}}, \bibinfo {author}
  {\bibfnamefont {P.}~\bibnamefont {T\"orm\"a}}, \ and\ \bibinfo {author}
  {\bibfnamefont {A.~W.}\ \bibnamefont {Sandvik}},\ }\href {\doibase
  10.1103/PhysRevLett.124.090502} {\bibfield  {journal} {\bibinfo  {journal}
  {Phys. Rev. Lett.}\ }\textbf {\bibinfo {volume} {124}},\ \bibinfo {pages}
  {090502} (\bibinfo {year} {2020})}\BibitemShut {NoStop}%
\bibitem [{\citenamefont {Kitai}\ \emph {et~al.}(2020)\citenamefont {Kitai},
  \citenamefont {Guo}, \citenamefont {Ju}, \citenamefont {Tanaka},
  \citenamefont {Tsuda}, \citenamefont {Shiomi},\ and\ \citenamefont
  {Tamura}}]{Kitai_2020}%
  \BibitemOpen
  \bibfield  {author} {\bibinfo {author} {\bibfnamefont {K.}~\bibnamefont
  {Kitai}}, \bibinfo {author} {\bibfnamefont {J.}~\bibnamefont {Guo}}, \bibinfo
  {author} {\bibfnamefont {S.}~\bibnamefont {Ju}}, \bibinfo {author}
  {\bibfnamefont {S.}~\bibnamefont {Tanaka}}, \bibinfo {author} {\bibfnamefont
  {K.}~\bibnamefont {Tsuda}}, \bibinfo {author} {\bibfnamefont
  {J.}~\bibnamefont {Shiomi}}, \ and\ \bibinfo {author} {\bibfnamefont
  {R.}~\bibnamefont {Tamura}},\ }\href {\doibase
  10.1103/PhysRevResearch.2.013319} {\bibfield  {journal} {\bibinfo  {journal}
  {Phys. Rev. Research}\ }\textbf {\bibinfo {volume} {2}},\ \bibinfo {pages}
  {013319} (\bibinfo {year} {2020})}\BibitemShut {NoStop}%
\bibitem [{\citenamefont {{Ide}}\ \emph {et~al.}(2020)\citenamefont {{Ide}},
  \citenamefont {{Asayama}}, \citenamefont {{Ueno}},\ and\ \citenamefont
  {{Ohzeki}}}]{SONY2020}%
  \BibitemOpen
  \bibfield  {author} {\bibinfo {author} {\bibfnamefont {N.}~\bibnamefont
  {{Ide}}}, \bibinfo {author} {\bibfnamefont {T.}~\bibnamefont {{Asayama}}},
  \bibinfo {author} {\bibfnamefont {H.}~\bibnamefont {{Ueno}}}, \ and\ \bibinfo
  {author} {\bibfnamefont {M.}~\bibnamefont {{Ohzeki}}},\ }in\ \href@noop {}
  {\emph {\bibinfo {booktitle} {2020 International Symposium on Information
  Theory and Its Applications (ISITA)}}}\ (\bibinfo {year} {2020})\ pp.\
  \bibinfo {pages} {91--95}\BibitemShut {NoStop}%
\bibitem [{\citenamefont {Suzuki}\ and\ \citenamefont
  {Okada}(2005)}]{suzuki_2005}%
  \BibitemOpen
  \bibfield  {author} {\bibinfo {author} {\bibfnamefont {S.}~\bibnamefont
  {Suzuki}}\ and\ \bibinfo {author} {\bibfnamefont {M.}~\bibnamefont {Okada}},\
  }\href@noop {} {\bibfield  {journal} {\bibinfo  {journal} {Journal of the
  Physical Society of Japan}\ }\textbf {\bibinfo {volume} {74}},\ \bibinfo
  {pages} {1649} (\bibinfo {year} {2005})}\BibitemShut {NoStop}%
\bibitem [{\citenamefont {J\"org}\ \emph {et~al.}(2008)\citenamefont {J\"org},
  \citenamefont {Krzakala}, \citenamefont {Kurchan},\ and\ \citenamefont
  {Maggs}}]{J_rg_2008}%
  \BibitemOpen
  \bibfield  {author} {\bibinfo {author} {\bibfnamefont {T.}~\bibnamefont
  {J\"org}}, \bibinfo {author} {\bibfnamefont {F.}~\bibnamefont {Krzakala}},
  \bibinfo {author} {\bibfnamefont {J.}~\bibnamefont {Kurchan}}, \ and\
  \bibinfo {author} {\bibfnamefont {A.~C.}\ \bibnamefont {Maggs}},\ }\href
  {\doibase 10.1103/PhysRevLett.101.147204} {\bibfield  {journal} {\bibinfo
  {journal} {Phys. Rev. Lett.}\ }\textbf {\bibinfo {volume} {101}},\ \bibinfo
  {pages} {147204} (\bibinfo {year} {2008})}\BibitemShut {NoStop}%
\bibitem [{\citenamefont {Young}\ \emph {et~al.}(2010)\citenamefont {Young},
  \citenamefont {Knysh},\ and\ \citenamefont {Smelyanskiy}}]{Young_2010}%
  \BibitemOpen
  \bibfield  {author} {\bibinfo {author} {\bibfnamefont {A.~P.}\ \bibnamefont
  {Young}}, \bibinfo {author} {\bibfnamefont {S.}~\bibnamefont {Knysh}}, \ and\
  \bibinfo {author} {\bibfnamefont {V.~N.}\ \bibnamefont {Smelyanskiy}},\
  }\href {\doibase 10.1103/PhysRevLett.104.020502} {\bibfield  {journal}
  {\bibinfo  {journal} {Phys. Rev. Lett.}\ }\textbf {\bibinfo {volume} {104}},\
  \bibinfo {pages} {020502} (\bibinfo {year} {2010})}\BibitemShut {NoStop}%
\bibitem [{\citenamefont {J{\"o}rg}\ \emph {et~al.}(2010)\citenamefont
  {J{\"o}rg}, \citenamefont {Krzakala}, \citenamefont {Kurchan}, \citenamefont
  {Maggs},\ and\ \citenamefont {Pujos}}]{J_rg_2010}%
  \BibitemOpen
  \bibfield  {author} {\bibinfo {author} {\bibfnamefont {T.}~\bibnamefont
  {J{\"o}rg}}, \bibinfo {author} {\bibfnamefont {F.}~\bibnamefont {Krzakala}},
  \bibinfo {author} {\bibfnamefont {J.}~\bibnamefont {Kurchan}}, \bibinfo
  {author} {\bibfnamefont {A.~C.}\ \bibnamefont {Maggs}}, \ and\ \bibinfo
  {author} {\bibfnamefont {J.}~\bibnamefont {Pujos}},\ }\href {\doibase
  10.1209/0295-5075/89/40004} {\bibfield  {journal} {\bibinfo  {journal} {{EPL}
  (Europhysics Letters)}\ }\textbf {\bibinfo {volume} {89}},\ \bibinfo {pages}
  {40004} (\bibinfo {year} {2010})}\BibitemShut {NoStop}%
\bibitem [{\citenamefont {J\"org}\ \emph {et~al.}(2010)\citenamefont {J\"org},
  \citenamefont {Krzakala}, \citenamefont {Semerjian},\ and\ \citenamefont
  {Zamponi}}]{J_rg_2010_prl}%
  \BibitemOpen
  \bibfield  {author} {\bibinfo {author} {\bibfnamefont {T.}~\bibnamefont
  {J\"org}}, \bibinfo {author} {\bibfnamefont {F.}~\bibnamefont {Krzakala}},
  \bibinfo {author} {\bibfnamefont {G.}~\bibnamefont {Semerjian}}, \ and\
  \bibinfo {author} {\bibfnamefont {F.}~\bibnamefont {Zamponi}},\ }\href
  {\doibase 10.1103/PhysRevLett.104.207206} {\bibfield  {journal} {\bibinfo
  {journal} {Phys. Rev. Lett.}\ }\textbf {\bibinfo {volume} {104}},\ \bibinfo
  {pages} {207206} (\bibinfo {year} {2010})}\BibitemShut {NoStop}%
\bibitem [{\citenamefont {Seki}\ and\ \citenamefont
  {Nishimori}(2012)}]{Seki_2012}%
  \BibitemOpen
  \bibfield  {author} {\bibinfo {author} {\bibfnamefont {Y.}~\bibnamefont
  {Seki}}\ and\ \bibinfo {author} {\bibfnamefont {H.}~\bibnamefont
  {Nishimori}},\ }\href {\doibase 10.1103/PhysRevE.85.051112} {\bibfield
  {journal} {\bibinfo  {journal} {Phys. Rev. E}\ }\textbf {\bibinfo {volume}
  {85}},\ \bibinfo {pages} {051112} (\bibinfo {year} {2012})}\BibitemShut
  {NoStop}%
\bibitem [{\citenamefont {Seki}\ and\ \citenamefont
  {Nishimori}(2015)}]{Seki_2015}%
  \BibitemOpen
  \bibfield  {author} {\bibinfo {author} {\bibfnamefont {Y.}~\bibnamefont
  {Seki}}\ and\ \bibinfo {author} {\bibfnamefont {H.}~\bibnamefont
  {Nishimori}},\ }\href {\doibase 10.1088/1751-8113/48/33/335301} {\bibfield
  {journal} {\bibinfo  {journal} {Journal of Physics A: Mathematical and
  Theoretical}\ }\textbf {\bibinfo {volume} {48}},\ \bibinfo {pages} {335301}
  (\bibinfo {year} {2015})}\BibitemShut {NoStop}%
\bibitem [{\citenamefont {Nishimori}\ and\ \citenamefont
  {Takada}(2017)}]{nishimori_2017}%
  \BibitemOpen
  \bibfield  {author} {\bibinfo {author} {\bibfnamefont {H.}~\bibnamefont
  {Nishimori}}\ and\ \bibinfo {author} {\bibfnamefont {K.}~\bibnamefont
  {Takada}},\ }\href {\doibase 10.3389/fict.2017.00002} {\bibfield  {journal}
  {\bibinfo  {journal} {Frontiers in ICT}\ }\textbf {\bibinfo {volume} {4}},\
  \bibinfo {pages} {2} (\bibinfo {year} {2017})}\BibitemShut {NoStop}%
\bibitem [{\citenamefont {Arai}\ \emph {et~al.}(2019)\citenamefont {Arai},
  \citenamefont {Ohzeki},\ and\ \citenamefont {Tanaka}}]{arai2019}%
  \BibitemOpen
  \bibfield  {author} {\bibinfo {author} {\bibfnamefont {S.}~\bibnamefont
  {Arai}}, \bibinfo {author} {\bibfnamefont {M.}~\bibnamefont {Ohzeki}}, \ and\
  \bibinfo {author} {\bibfnamefont {K.}~\bibnamefont {Tanaka}},\ }\href
  {\doibase 10.1103/PhysRevE.99.032120} {\bibfield  {journal} {\bibinfo
  {journal} {Phys. Rev. E}\ }\textbf {\bibinfo {volume} {99}},\ \bibinfo
  {pages} {032120} (\bibinfo {year} {2019})}\BibitemShut {NoStop}%
\bibitem [{\citenamefont {Susa}\ \emph
  {et~al.}(2018{\natexlab{a}})\citenamefont {Susa}, \citenamefont {Yamashiro},
  \citenamefont {Yamamoto},\ and\ \citenamefont {Nishimori}}]{Susa_2018a}%
  \BibitemOpen
  \bibfield  {author} {\bibinfo {author} {\bibfnamefont {Y.}~\bibnamefont
  {Susa}}, \bibinfo {author} {\bibfnamefont {Y.}~\bibnamefont {Yamashiro}},
  \bibinfo {author} {\bibfnamefont {M.}~\bibnamefont {Yamamoto}}, \ and\
  \bibinfo {author} {\bibfnamefont {H.}~\bibnamefont {Nishimori}},\ }\href
  {\doibase 10.7566/JPSJ.87.023002} {\bibfield  {journal} {\bibinfo  {journal}
  {Journal of the Physical Society of Japan}\ }\textbf {\bibinfo {volume}
  {87}},\ \bibinfo {pages} {023002} (\bibinfo {year} {2018}{\natexlab{a}})},\
  \Eprint {http://arxiv.org/abs/https://doi.org/10.7566/JPSJ.87.023002}
  {https://doi.org/10.7566/JPSJ.87.023002} \BibitemShut {NoStop}%
\bibitem [{\citenamefont {Susa}\ \emph
  {et~al.}(2018{\natexlab{b}})\citenamefont {Susa}, \citenamefont {Yamashiro},
  \citenamefont {Yamamoto}, \citenamefont {Hen}, \citenamefont {Lidar},\ and\
  \citenamefont {Nishimori}}]{Susa_2018b}%
  \BibitemOpen
  \bibfield  {author} {\bibinfo {author} {\bibfnamefont {Y.}~\bibnamefont
  {Susa}}, \bibinfo {author} {\bibfnamefont {Y.}~\bibnamefont {Yamashiro}},
  \bibinfo {author} {\bibfnamefont {M.}~\bibnamefont {Yamamoto}}, \bibinfo
  {author} {\bibfnamefont {I.}~\bibnamefont {Hen}}, \bibinfo {author}
  {\bibfnamefont {D.~A.}\ \bibnamefont {Lidar}}, \ and\ \bibinfo {author}
  {\bibfnamefont {H.}~\bibnamefont {Nishimori}},\ }\href {\doibase
  10.1103/PhysRevA.98.042326} {\bibfield  {journal} {\bibinfo  {journal} {Phys.
  Rev. A}\ }\textbf {\bibinfo {volume} {98}},\ \bibinfo {pages} {042326}
  (\bibinfo {year} {2018}{\natexlab{b}})}\BibitemShut {NoStop}%
\bibitem [{\citenamefont {Perdomo-Ortiz}\ \emph {et~al.}(2011)\citenamefont
  {Perdomo-Ortiz}, \citenamefont {Venegas-Andraca},\ and\ \citenamefont
  {Aspuru-Guzik}}]{Perdomo2011}%
  \BibitemOpen
  \bibfield  {author} {\bibinfo {author} {\bibfnamefont {A.}~\bibnamefont
  {Perdomo-Ortiz}}, \bibinfo {author} {\bibfnamefont {S.~E.}\ \bibnamefont
  {Venegas-Andraca}}, \ and\ \bibinfo {author} {\bibfnamefont {A.}~\bibnamefont
  {Aspuru-Guzik}},\ }\href {\doibase 10.1007/s11128-010-0168-z} {\bibfield
  {journal} {\bibinfo  {journal} {Quantum Information Processing}\ }\textbf
  {\bibinfo {volume} {10}},\ \bibinfo {pages} {33} (\bibinfo {year}
  {2011})}\BibitemShut {NoStop}%
\bibitem [{\citenamefont {Chancellor}(2017)}]{Chancellor_2017}%
  \BibitemOpen
  \bibfield  {author} {\bibinfo {author} {\bibfnamefont {N.}~\bibnamefont
  {Chancellor}},\ }\href {\doibase 10.1088/1367-2630/aa59c4} {\bibfield
  {journal} {\bibinfo  {journal} {New Journal of Physics}\ }\textbf {\bibinfo
  {volume} {19}},\ \bibinfo {pages} {023024} (\bibinfo {year}
  {2017})}\BibitemShut {NoStop}%
\bibitem [{\citenamefont {Ohkuwa}\ \emph {et~al.}(2018)\citenamefont {Ohkuwa},
  \citenamefont {Nishimori},\ and\ \citenamefont {Lidar}}]{ohkuwa_2018}%
  \BibitemOpen
  \bibfield  {author} {\bibinfo {author} {\bibfnamefont {M.}~\bibnamefont
  {Ohkuwa}}, \bibinfo {author} {\bibfnamefont {H.}~\bibnamefont {Nishimori}}, \
  and\ \bibinfo {author} {\bibfnamefont {D.~A.}\ \bibnamefont {Lidar}},\ }\href
  {\doibase 10.1103/PhysRevA.98.022314} {\bibfield  {journal} {\bibinfo
  {journal} {Phys. Rev. A}\ }\textbf {\bibinfo {volume} {98}},\ \bibinfo
  {pages} {022314} (\bibinfo {year} {2018})}\BibitemShut {NoStop}%
\bibitem [{\citenamefont {Yamashiro}\ \emph {et~al.}(2019)\citenamefont
  {Yamashiro}, \citenamefont {Ohkuwa}, \citenamefont {Nishimori},\ and\
  \citenamefont {Lidar}}]{Yamashiro_2019}%
  \BibitemOpen
  \bibfield  {author} {\bibinfo {author} {\bibfnamefont {Y.}~\bibnamefont
  {Yamashiro}}, \bibinfo {author} {\bibfnamefont {M.}~\bibnamefont {Ohkuwa}},
  \bibinfo {author} {\bibfnamefont {H.}~\bibnamefont {Nishimori}}, \ and\
  \bibinfo {author} {\bibfnamefont {D.~A.}\ \bibnamefont {Lidar}},\ }\href
  {\doibase 10.1103/PhysRevA.100.052321} {\bibfield  {journal} {\bibinfo
  {journal} {Phys. Rev. A}\ }\textbf {\bibinfo {volume} {100}},\ \bibinfo
  {pages} {052321} (\bibinfo {year} {2019})}\BibitemShut {NoStop}%
\bibitem [{\citenamefont {Passarelli}\ \emph {et~al.}(2020)\citenamefont
  {Passarelli}, \citenamefont {Yip}, \citenamefont {Lidar}, \citenamefont
  {Nishimori},\ and\ \citenamefont {Lucignano}}]{Passarelli_2020}%
  \BibitemOpen
  \bibfield  {author} {\bibinfo {author} {\bibfnamefont {G.}~\bibnamefont
  {Passarelli}}, \bibinfo {author} {\bibfnamefont {K.-W.}\ \bibnamefont {Yip}},
  \bibinfo {author} {\bibfnamefont {D.~A.}\ \bibnamefont {Lidar}}, \bibinfo
  {author} {\bibfnamefont {H.}~\bibnamefont {Nishimori}}, \ and\ \bibinfo
  {author} {\bibfnamefont {P.}~\bibnamefont {Lucignano}},\ }\href {\doibase
  10.1103/PhysRevA.101.022331} {\bibfield  {journal} {\bibinfo  {journal}
  {Phys. Rev. A}\ }\textbf {\bibinfo {volume} {101}},\ \bibinfo {pages}
  {022331} (\bibinfo {year} {2020})}\BibitemShut {NoStop}%
\bibitem [{Note1()}]{Note1}%
  \BibitemOpen
  \bibinfo {note} {Originally, the ARA is applied when we search the ground
  state in the target Hamiltonian. The search of the ground state corresponds
  to the maximum a posteriori (MAP) estimation. The main problem of the MPM
  estimation is not to search the ground state but to sample the low energy
  state from the Gibbs--Boltzmann distribution. Strictly speaking, we should
  not utilize the term ``annealing'' because we do not perform that in this
  paper. Since the MPM estimation in the zero-temperature limit corresponds to
  the MAP estimation, we utilize the term ``ARA''.}\BibitemShut {Stop}%
\bibitem [{\citenamefont {Amin}(2015)}]{Amin2015}%
  \BibitemOpen
  \bibfield  {author} {\bibinfo {author} {\bibfnamefont {M.~H.}\ \bibnamefont
  {Amin}},\ }\href {\doibase 10.1103/PhysRevA.92.052323} {\bibfield  {journal}
  {\bibinfo  {journal} {Phys. Rev. A}\ }\textbf {\bibinfo {volume} {92}},\
  \bibinfo {pages} {052323} (\bibinfo {year} {2015})}\BibitemShut {NoStop}%
\bibitem [{\citenamefont {Chancellor}\ \emph {et~al.}(2016)\citenamefont
  {Chancellor}, \citenamefont {Szoke}, \citenamefont {Vinci}, \citenamefont
  {Aeppli},\ and\ \citenamefont {Warburton}}]{Chancellor2016}%
  \BibitemOpen
  \bibfield  {author} {\bibinfo {author} {\bibfnamefont {N.}~\bibnamefont
  {Chancellor}}, \bibinfo {author} {\bibfnamefont {S.}~\bibnamefont {Szoke}},
  \bibinfo {author} {\bibfnamefont {W.}~\bibnamefont {Vinci}}, \bibinfo
  {author} {\bibfnamefont {G.}~\bibnamefont {Aeppli}}, \ and\ \bibinfo {author}
  {\bibfnamefont {P.~A.}\ \bibnamefont {Warburton}},\ }\href {\doibase
  10.1038/srep22318} {\bibfield  {journal} {\bibinfo  {journal} {Scientific
  Reports}\ }\textbf {\bibinfo {volume} {6}},\ \bibinfo {pages} {22318}
  (\bibinfo {year} {2016})}\BibitemShut {NoStop}%
\bibitem [{\citenamefont {Nishimori}(1993)}]{nishimori_1993}%
  \BibitemOpen
  \bibfield  {author} {\bibinfo {author} {\bibfnamefont {H.}~\bibnamefont
  {Nishimori}},\ }\href {\doibase 10.1143/JPSJ.62.2973} {\bibfield  {journal}
  {\bibinfo  {journal} {Journal of the Physical Society of Japan}\ }\textbf
  {\bibinfo {volume} {62}},\ \bibinfo {pages} {2973} (\bibinfo {year}
  {1993})}\BibitemShut {NoStop}%
\bibitem [{\citenamefont {Suzuki}(1976)}]{suzuki_1976}%
  \BibitemOpen
  \bibfield  {author} {\bibinfo {author} {\bibfnamefont {M.}~\bibnamefont
  {Suzuki}},\ }\href {\doibase 10.1007/BF01609348} {\bibfield  {journal}
  {\bibinfo  {journal} {Communications in Mathematical Physics}\ }\textbf
  {\bibinfo {volume} {51}},\ \bibinfo {pages} {183} (\bibinfo {year}
  {1976})}\BibitemShut {NoStop}%
\bibitem [{\citenamefont {Sherrington}\ and\ \citenamefont
  {Kirkpatrick}(1975)}]{replica_method}%
  \BibitemOpen
  \bibfield  {author} {\bibinfo {author} {\bibfnamefont {D.}~\bibnamefont
  {Sherrington}}\ and\ \bibinfo {author} {\bibfnamefont {S.}~\bibnamefont
  {Kirkpatrick}},\ }\href {\doibase 10.1103/PhysRevLett.35.1792} {\bibfield
  {journal} {\bibinfo  {journal} {Phys. Rev. Lett.}\ }\textbf {\bibinfo
  {volume} {35}},\ \bibinfo {pages} {1792} (\bibinfo {year}
  {1975})}\BibitemShut {NoStop}%
\bibitem [{sup()}]{supp1}%
  \BibitemOpen
  \bibinfo {note} {See Supplemental Material material}\BibitemShut {NoStop}%
\bibitem [{\citenamefont {{Yoshida}}\ and\ \citenamefont
  {{Tanaka}}(2006)}]{Yoshida_2006}%
  \BibitemOpen
  \bibfield  {author} {\bibinfo {author} {\bibfnamefont {M.}~\bibnamefont
  {{Yoshida}}}\ and\ \bibinfo {author} {\bibfnamefont {T.}~\bibnamefont
  {{Tanaka}}},\ }in\ \href {\doibase 10.1109/ISIT.2006.262014} {\emph {\bibinfo
  {booktitle} {2006 IEEE International Symposium on Information Theory}}}\
  (\bibinfo {year} {2006})\ pp.\ \bibinfo {pages} {2378--2382}\BibitemShut
  {NoStop}%
\bibitem [{Note2()}]{Note2}%
  \BibitemOpen
  \bibinfo {note} {We verify the dependence of the order parameters on the
  Trotter number in our simulations. As we increase the Trotter number, the
  deviation of the correlation between the Trotter slices from the RS solutions
  decreases. The qualitative results are similar to those in the
  magnetization.}\BibitemShut {Stop}%
\bibitem [{\citenamefont {Opper}\ and\ \citenamefont {Saad}(2001)}]{opperbook}%
  \BibitemOpen
  \bibinfo {editor} {\bibfnamefont {M.}~\bibnamefont {Opper}}\ and\ \bibinfo
  {editor} {\bibfnamefont {D.}~\bibnamefont {Saad}},\ eds.,\ \href@noop {}
  {\emph {\bibinfo {title} {Advanced mean field methods: theory and
  practice}}},\ Neural Information Processing\ (\bibinfo  {publisher} {MIT},\
  \bibinfo {year} {2001})\BibitemShut {NoStop}%
\bibitem [{\citenamefont {Kabashima}(2003)}]{Kabashima_2003}%
  \BibitemOpen
  \bibfield  {author} {\bibinfo {author} {\bibfnamefont {Y.}~\bibnamefont
  {Kabashima}},\ }\href {\doibase 10.1088/0305-4470/36/43/030} {\bibfield
  {journal} {\bibinfo  {journal} {Journal of Physics A: Mathematical and
  General}\ }\textbf {\bibinfo {volume} {36}},\ \bibinfo {pages} {11111}
  (\bibinfo {year} {2003})}\BibitemShut {NoStop}%
\bibitem [{\citenamefont {Donoho}\ \emph {et~al.}(2009)\citenamefont {Donoho},
  \citenamefont {Maleki},\ and\ \citenamefont {Montanari}}]{Donoho_2009b}%
  \BibitemOpen
  \bibfield  {author} {\bibinfo {author} {\bibfnamefont {D.~L.}\ \bibnamefont
  {Donoho}}, \bibinfo {author} {\bibfnamefont {A.}~\bibnamefont {Maleki}}, \
  and\ \bibinfo {author} {\bibfnamefont {A.}~\bibnamefont {Montanari}},\ }\href
  {\doibase 10.1073/pnas.0909892106} {\bibfield  {journal} {\bibinfo  {journal}
  {Proceedings of the National Academy of Sciences}\ }\textbf {\bibinfo
  {volume} {106}},\ \bibinfo {pages} {18914} (\bibinfo {year}
  {2009})}\BibitemShut {NoStop}%
\bibitem [{Note3()}]{Note3}%
  \BibitemOpen
  \bibinfo {note} {\protect \url
  {https://github.com/OpenJij/OpenJij}}\BibitemShut {NoStop}%
\bibitem [{\citenamefont {{Silvio Franz}}\ and\ \citenamefont {{Giorgio
  Parisi}}(1995)}]{Franz1995}%
  \BibitemOpen
  \bibfield  {author} {\bibinfo {author} {\bibnamefont {{Silvio Franz}}}\ and\
  \bibinfo {author} {\bibnamefont {{Giorgio Parisi}}},\ }\href {\doibase
  10.1051/jp1:1995201} {\bibfield  {journal} {\bibinfo  {journal} {J. Phys. I
  France}\ }\textbf {\bibinfo {volume} {5}},\ \bibinfo {pages} {1401} (\bibinfo
  {year} {1995})}\BibitemShut {NoStop}%
\bibitem [{\citenamefont {Franz}\ and\ \citenamefont
  {Parisi}(1997)}]{Franz1997}%
  \BibitemOpen
  \bibfield  {author} {\bibinfo {author} {\bibfnamefont {S.}~\bibnamefont
  {Franz}}\ and\ \bibinfo {author} {\bibfnamefont {G.}~\bibnamefont {Parisi}},\
  }\href {\doibase 10.1103/PhysRevLett.79.2486} {\bibfield  {journal} {\bibinfo
   {journal} {Phys. Rev. Lett.}\ }\textbf {\bibinfo {volume} {79}},\ \bibinfo
  {pages} {2486} (\bibinfo {year} {1997})}\BibitemShut {NoStop}%
\bibitem [{\citenamefont {Franz}\ and\ \citenamefont
  {Parisi}(1998)}]{Franz1998}%
  \BibitemOpen
  \bibfield  {author} {\bibinfo {author} {\bibfnamefont {S.}~\bibnamefont
  {Franz}}\ and\ \bibinfo {author} {\bibfnamefont {G.}~\bibnamefont {Parisi}},\
  }\href {\doibase https://doi.org/10.1016/S0378-4371(98)00315-X} {\bibfield
  {journal} {\bibinfo  {journal} {Physica A: Statistical Mechanics and its
  Applications}\ }\textbf {\bibinfo {volume} {261}},\ \bibinfo {pages} {317 }
  (\bibinfo {year} {1998})}\BibitemShut {NoStop}%
\bibitem [{\citenamefont {Huang}\ \emph {et~al.}(2013)\citenamefont {Huang},
  \citenamefont {Wong},\ and\ \citenamefont {Kabashima}}]{Huang2013}%
  \BibitemOpen
  \bibfield  {author} {\bibinfo {author} {\bibfnamefont {H.}~\bibnamefont
  {Huang}}, \bibinfo {author} {\bibfnamefont {K.~Y.~M.}\ \bibnamefont {Wong}},
  \ and\ \bibinfo {author} {\bibfnamefont {Y.}~\bibnamefont {Kabashima}},\
  }\href {\doibase 10.1088/1751-8113/46/37/375002} {\bibfield  {journal}
  {\bibinfo  {journal} {Journal of Physics A: Mathematical and Theoretical}\
  }\textbf {\bibinfo {volume} {46}},\ \bibinfo {pages} {375002} (\bibinfo
  {year} {2013})}\BibitemShut {NoStop}%
\bibitem [{\citenamefont {Huang}\ and\ \citenamefont
  {Kabashima}(2014)}]{Huang2014}%
  \BibitemOpen
  \bibfield  {author} {\bibinfo {author} {\bibfnamefont {H.}~\bibnamefont
  {Huang}}\ and\ \bibinfo {author} {\bibfnamefont {Y.}~\bibnamefont
  {Kabashima}},\ }\href {\doibase 10.1103/PhysRevE.90.052813} {\bibfield
  {journal} {\bibinfo  {journal} {Phys. Rev. E}\ }\textbf {\bibinfo {volume}
  {90}},\ \bibinfo {pages} {052813} (\bibinfo {year} {2014})}\BibitemShut
  {NoStop}%
\bibitem [{\citenamefont {de~Almeida}\ and\ \citenamefont
  {Thouless}(1978)}]{Almeida_1978}%
  \BibitemOpen
  \bibfield  {author} {\bibinfo {author} {\bibfnamefont {J.~R.~L.}\
  \bibnamefont {de~Almeida}}\ and\ \bibinfo {author} {\bibfnamefont {D.~J.}\
  \bibnamefont {Thouless}},\ }\href {\doibase 10.1088/0305-4470/11/5/028}
  {\bibfield  {journal} {\bibinfo  {journal} {Journal of Physics A:
  Mathematical and General}\ }\textbf {\bibinfo {volume} {11}},\ \bibinfo
  {pages} {983} (\bibinfo {year} {1978})}\BibitemShut {NoStop}%
\bibitem [{\citenamefont {Kabashima}(2008)}]{Kabashima_2008}%
  \BibitemOpen
  \bibfield  {author} {\bibinfo {author} {\bibfnamefont {Y.}~\bibnamefont
  {Kabashima}},\ }\href {\doibase 10.1088/1742-6596/95/1/012001} {\bibfield
  {journal} {\bibinfo  {journal} {Journal of Physics: Conference Series}\
  }\textbf {\bibinfo {volume} {95}},\ \bibinfo {pages} {012001} (\bibinfo
  {year} {2008})}\BibitemShut {NoStop}%
\bibitem [{\citenamefont {Sakata}\ and\ \citenamefont
  {Xu}(2018)}]{Sakata_2018}%
  \BibitemOpen
  \bibfield  {author} {\bibinfo {author} {\bibfnamefont {A.}~\bibnamefont
  {Sakata}}\ and\ \bibinfo {author} {\bibfnamefont {Y.}~\bibnamefont {Xu}},\
  }\href {\doibase 10.1088/1742-5468/aab051} {\bibfield  {journal} {\bibinfo
  {journal} {Journal of Statistical Mechanics: Theory and Experiment}\ }\textbf
  {\bibinfo {volume} {2018}},\ \bibinfo {pages} {033404} (\bibinfo {year}
  {2018})}\BibitemShut {NoStop}%
\bibitem [{\citenamefont {{Krauth, Werner}}\ and\ \citenamefont {{M\'ezard,
  Marc}}(1989)}]{Krauth_1989}%
  \BibitemOpen
  \bibfield  {author} {\bibinfo {author} {\bibnamefont {{Krauth, Werner}}}\
  and\ \bibinfo {author} {\bibnamefont {{M\'ezard, Marc}}},\ }\href {\doibase
  10.1051/jphys:0198900500200305700} {\bibfield  {journal} {\bibinfo  {journal}
  {J. Phys. France}\ }\textbf {\bibinfo {volume} {50}},\ \bibinfo {pages}
  {3057} (\bibinfo {year} {1989})}\BibitemShut {NoStop}%
\end{thebibliography}%
\end{document}